\documentclass[openacc]{rsproca_new}
\usepackage{amsmath,amssymb,mathrsfs,graphicx}
\usepackage[english]{babel}
\begin{document}

\title{\boldmath Neutron reflectometry with the Multi-Blade 10B-based detector}

\author{G. Mauri$^{1,2}$, F. Messi$^{1,3}$, M. Anastasopoulos$^{1}$, T. Arnold$^{1}$, A. Glavic$^{4}$, C. H\"{o}glund$^{1,5}$, T. Ilves$^{3}$, I. Lopez Higuera$^{1}$, P. Pazmandi$^{6}$, D. Raspino$^{7}$, L. Robinson$^{1}$, S. Schmidt$^{1,8}$, P. Svensson$^{1}$, D. Varga$^{6}$, R. Hall-Wilton$^{1,9}$, F. Piscitelli$^{1}$}

\address{$^{1}$European Spallation Source ERIC, P.O. Box 176, SE-221 00 Lund, SE.\\
$^{2}$Department of Physics, University of Perugia, Piazza Universit\`a 1, 06123 Perugia, IT. \\
$^{3}$Division of Nuclear Physics, Lund University, \\P.O. Box 118, SE-22100 Lund, Sweden. \\
$^{4}$Laboratory for Neutron Scattering and Imaging, Paul Scherrer Institute, 5232 Villigen PSI, Switzerland. \\
$^{5}$Department of Physics, Chemistry and Biology, Link\"{o}ping University, SE-581 83 Link\"{o}ping, Sweden. \\
$^{6}$Wigner Research Centre for Physics, \\Konkoly Thege Mikl\'os \'ut 29-33, H-1121 Budapest, Hungary. \\
$^{7}$ISIS Neutron and Muon Source, Harwell Oxford, Didcot OX11 0QX, United Kingdom.\\
$^{8}$IHI Ionbond AG, Industriestrasse 211, 4600 Olten, Switzerland.\\
$^{9}$Mid-Sweden University, SE-851 70 Sundsvall, SE.}

%\subject{xxxxx, xxxxx, xxxx}


\corres{F. Piscitelli\\
\email{francesco.piscitelli@esss.se}}
 
\begin{abstract}
The Multi-Blade is a Boron-10-based gaseous detector developed for neutron reflectometry instruments at the European Spallation Source (ESS) in Sweden. The main challenges for neutron reflectometry detectors are the instantaneous counting rate and spatial resolution. The Multi-Blade has been tested on the CRISP reflectometer at the ISIS neutron and muon source in UK. A campaign of scientific measurements has been performed to study the Multi-Blade response in real instrumental conditions. The results of these tests are discussed in this manuscript.
\end{abstract}
\maketitle
\section{Introduction}\label{intro}
The Multi-Blade~\cite{MIO_MB2017,MIO_MB2014,MIO_MBproc,MIO_MyThesis,MIO_MB16CRISP_jinst} is a $\mathrm{^{10}B}$-based detector for neutron reflectometry instruments~\cite{INSTR_D17,INSTR_FIGARO, INSTR_ISIS_R}.The detector requirements are set by the two reflectometers that are being designed for the European Spallation Source (ESS~\cite{ESS}) in Sweden: FREIA~\cite{INSTR_FREIA,INSTR_FREIA2} (horizontal reflectometer) and ESTIA~\cite{INSTR_ESTIA,INSTR_ESTIA1,INSTR_ESTIA2} (vertical reflectometer). In the past few years several methods have been proposed to improve the performance of reflectometry instruments and the ESS reflectometers are based on these new concepts. 

Neutron reflectometry and off-specular scattering are powerful techniques to study depth profiles and in-plane correlations of thin film samples~\cite{MISC_pike2002scattering,R_fermi,Lauter2016}. In a typical neutron reflection experiment the reflection of neutrons as a function of the wave-vector transfer in direction of the surface normal, $q_z$, is measured:
\begin{equation}
\label{eqaf1}
q_z=\frac{4\pi}{\lambda}\sin(\theta)
\end{equation}
where $\lambda$ is the neutron wavelength and $\theta$ is the angle between the beam and sample surface (which is the same for incident and reflected beam, $\alpha_i=\alpha_f=\theta$). 
\\ Neutron reflection follows the same fundamental equations as optical reflectivity but with different refractive indices. The optical properties of neutron propagation arise from the fact that quantum-mechanically the neutron is described by a wave-function. The potential ($V$) in the Schr\"{o}dinger equation, which is the averaged density of the scattering lengths of the material, plays the role of a refractive index.
\begin{equation}\label{eqasld}
V=\frac{2\pi\hbar^2}{m_n} N_b=\frac{2\pi\hbar^2}{m_n}\sum_i b_i n_i
\end{equation}
where $m_n$ the neutron mass, $\hbar$ is the Plank's constant, $N_b$ is the \emph{scattering length density} of the medium, where $n_i$ is the number of nuclei per unit volume and $b_i$ is the coherent scattering length of nucleus $i$, because we take the spin-average (non-polarized beam or sample).
\\ The neutron refractive index is given by the scattering length density of its constituent nuclei and the neutron wavelength. As with light, total reflection may occur when neutrons pass from a medium of higher refractive index to one of lower refractive index. The angle where no neutrons penetrate the surface, hence all of them are reflected, is called \emph{critical angle} (or equivalently \emph{critical edge}): the reflectivity of neutrons of a given wavelength (or given $q$) from a bulk interface is unity at smaller angles and falls sharply at larger angles. As with light, interference can occur between waves reflected at the top and at the bottom of a thin film, which gives rise to interference fringes in the reflectivity profile~\cite{MISC_pike2002scattering}.
\\The typical neutron wavelengths ($\lambda$) in a reflectometry experiment are in the range of 2 - 20 \AA , which corresponds to a range between 0.05 and 3 nm$^{-1}$ in the wave-vector transfer ($q_z$). In the real space this corresponds to length-scales between 2 and 100 nm~\cite{R_offspec0_Ott}. The limits are imposed both by the measurement range and the instrumental resolution.
In the case of off-specular scattering it is possible to investigate objects in the plane with a correlation length of the order of several micrometers (50 to 0.5 $\mu$m). The upper limit is set by the resolution of the intruments and the size of the direct beam. The lower limit is determined by the available neutron flux~\cite{R_offspec0_Ott}. 
\\ In the last two decades the reflectometers have been optimized and allow to measure reflectivities below $10^{-6}$, enough for most experiments~\cite{OTT_general}. The next step is to increase the available flux, this leads to a significant speed up of reflectivity measurements and the possibility of using smaller samples. 
\\Several techniques have been recently proposed to improve the operating performance of reflectometry instruments. The methods are based on spin-space~\cite{INSTR_R_Spin2}, time-space~\cite{OTT_tiltof} or energy-space encoding~\cite{OTT_gradtof,OTT_refocus,R_Cubitt1,R_Cubitt2}. The first technique is used for off-specular measurements~\cite{INSTR_R_Spin} and encodes the incident angle by the rotation of the neutron spin in a magnetic field. The time-space encoding (TilTOF) enables an increase in the incoming flux on the sample, removing the chopper and modulating mechanically the angle of the sample to determine the time shape of the beam, and thus the wavelength. The idea of energy-space encoding is to analyse the neutron energies through a spatial spread of the reflected beam produced by an energy dispersive device, either a refractive crystal~\cite{R_Cubitt1,R_Cubitt2} or a magnetic field gradient~\cite{OTT_gradtof}. It is also possible to correlate the neutron wavelength and the incident angle, hence before the sample, using a divergent beam focused on the sample. The REFocus~\cite{OTT_refocus} technique employs an elliptical graded multilayer monochromator to focus the neutrons on the sample.
This concept has been modified and adapted to the time-of-flight instrument AMOR at PSI~\cite{PSI}, using an elliptic-shaped reflector: the \textit{Selene} guide~\cite{INSTR_ESTIA0,INSTR_ESTIA1}. A scaled-down demonstrator is implemented on AMOR at PSI~\cite{INSTR_ESTIA2} to prove the concept and to test the performances of the guide. The full-scale Selene guide will be a primary feature for ESTIA a reflectometer instrument at European Spallation Source (ESS, Lund, Sweden) now under construction.
\\ The general aim of all these optimizations is to increase the available neutron flux on the sample; thus time resolved measurements for kinetic studies can be performed, smaller samples can be used, faster measurements scaling down from hours, typical time for present day reflectivity experiments, to minutes can be performed. This gives the possibility to probe a dynamic range of reflectivity measurements down to $10^{-7}$.
\\ These improvements represent a challenge not only for the instrument design, but also for the performance of the detector technologies to be employed. The current detector technology is reaching fundamental limits, e.g. a sub-mm spatial resolution (Full-Width-Half-Maximum, FWHM) and high counting rate capabilities are required for the new instruments and it is not achievable with the state-of-the-art technology. The expected instantaneous local flux at the detector for the reflectometers at ESS is about $10^{5}/s/mm^2$~\cite{ESS_TDR,DET_rates,HE3S_kirstein}. Note that the current detector technology is already limiting the performance of the neutron reflectometers at existing sources (pulsed and reactors).

The Multi-Blade detector has been designed to fulfill these challenging requirements in terms of spatial resolution and counting rate capability. A demonstrator has been installed at the neutron reflectometer CRISP~\cite{CRISP1} at the ISIS neutron and muon source in UK~\cite{ISIS}. The detector has been characterized and a series of scientific measurements with several samples have been performed. The technical characterization of the Multi-Blade is not treated in this manuscript and a detailed description can be found in~ \cite{MIO_MB16CRISP_jinst}.
The performance of the Multi-Blade detector concerning the scientific measurements is subject of this paper. The aim of this test is not only to prove the capabilities of the detector in an actual instrument, but to show as well the improvements that arise from operating the CRISP reflectometer in a configuration which reproduces the ESTIA operation mode. This is exclusively possible by exploiting the features of the Multi-Blade.

\section{The Multi-Blade detector tested at CRISP}

\begin{figure}[htbp]
\centering
\includegraphics[width=.8\textwidth,keepaspectratio]{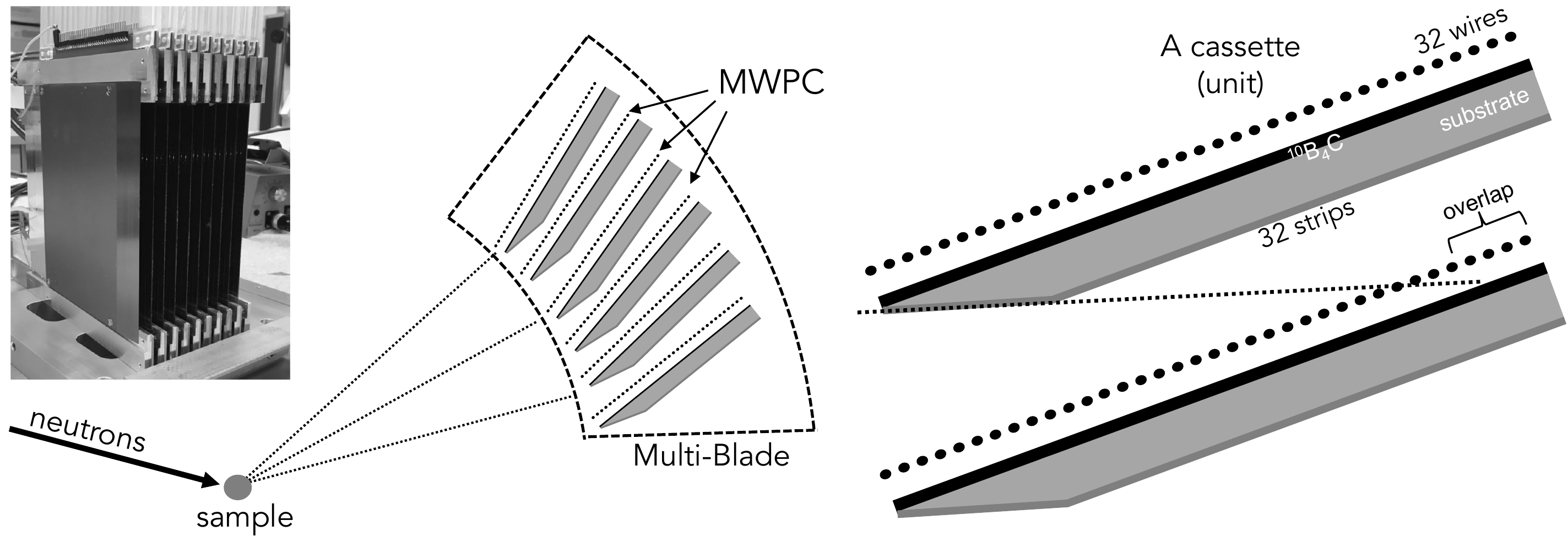}
\caption{\label{fig99} \footnotesize Schematic view of the cross-section of the Multi-Blade detector made up of identical units (cassettes) arranged adjacent to each other. Each cassette holds a $\mathrm{^{10}B_4C}$-layer; the readout is performed through a plane of wires and a plane of strips.}
\end{figure}

The Multi-Blade is a stack of Multi Wire Proportional Chambers (MWPC) operated at atmospheric pressure with a continuous gas flow ($\mathrm{Ar/CO_2}$ 80/20 mixture by volume). A sketch of the Multi-Blade detector is shown in figure~\ref{fig99}. The Multi-Blade is made up of identical units, the so-called `cassettes'. Each cassette holds a `blade' (a flat substrate coated with $\mathrm{^{10}B_4C}$~\cite{B4C_carina,B4C_carina3,B4C_Schmidt}) and a two-dimensional readout system, which consists of a plane of 32 wires and a plane of 32 strips. Each $\mathrm{^{10}B_4C}$-converter (blade) is inclined at grazing angle ($\theta = 5$ degrees) with respect to the incoming neutron beam. The cassettes are arranged over a circle around the sample and they have some overlap; i.e. each blade makes a shadow over the adjacent in order to avoid dead areas. The detailed description of the detector can be found in~\cite{MIO_MB2017, MIO_MB16CRISP_jinst}
\\The present detector consists of 9 units (576 channels in total). Each channel (64 per cassette) is readout individually, connected to a FET-based charge pre-amplifier and shaping amplifier. Each 32-channel board is connected to a CAEN V1740D digitizer ($12\,$bit, $62.5\,$MS/s)~\cite{EL_CAEN}. There are 6 digitizers in total and each can readout 64 channels , i.e. one cassette. Thus, out of the 9 cassettes, 6 could be used simultaneously in the tests. The 6 digitizers can be synchronized to the same clock source and a TTL logic signal can be sent to one of them and propagated to reset the time-stamp which is associated to an event. This feature is needed to perform any type of Time-of-Flight (ToF) measurement. In the case of CRISP, the reset of the time-stamp is given by the proton pulse of the ISIS source.
\\ The raw data from the read-out electronic system is reduced to a triplet $(X,Y,ToF)$ which identifies a single neutron event. The reconstruction algorithm used is described in detail in~\cite{MIO_MB16CRISP_jinst}. The triplets define a three-dimensional space containing the information where in the detector the neutron was detected with associated ToF. 
The spatial coordinates, X and Y, of a triplet, reflect the physical channels in the detector (32 wires and 32 strips) projected over the detector entrance window ( i.e. the projection of the blades toward the sample position). The Multi-Blade detector is, indeed, a three-dimensional detector, but the depth coordinate (Z) is integrated over. 

\section{Experimental setup on CRISP}\label{rifle}

CRISP is an horizontal neutron reflectometer at ISIS, Target Station 1, that uses a broad band neutron Time-of-Flight (ToF) method for determining the wavelength, (and hence $q$), at fixed angles ($\theta$). A detailed description of the CRISP reflectometer can be found in~\cite{CRISP1}. The instrument views an hydrogen moderator giving an effective wavelength range of $0.5-6.5$\AA\, at the source frequency of $50\,$Hz. The wavelength band extends up to 13\,\AA\, if operated at $25\,$Hz. A frame overlap mirror suppresses the wavelengths above 13\,\AA. The distance from the moderator to the sample is $10.25\,$m and the sample to the Multi-Blade detector distance is approximately $2.3\,$m. 
\begin{figure}[htbp]
\centering
\includegraphics[width=0.8\textwidth,keepaspectratio]{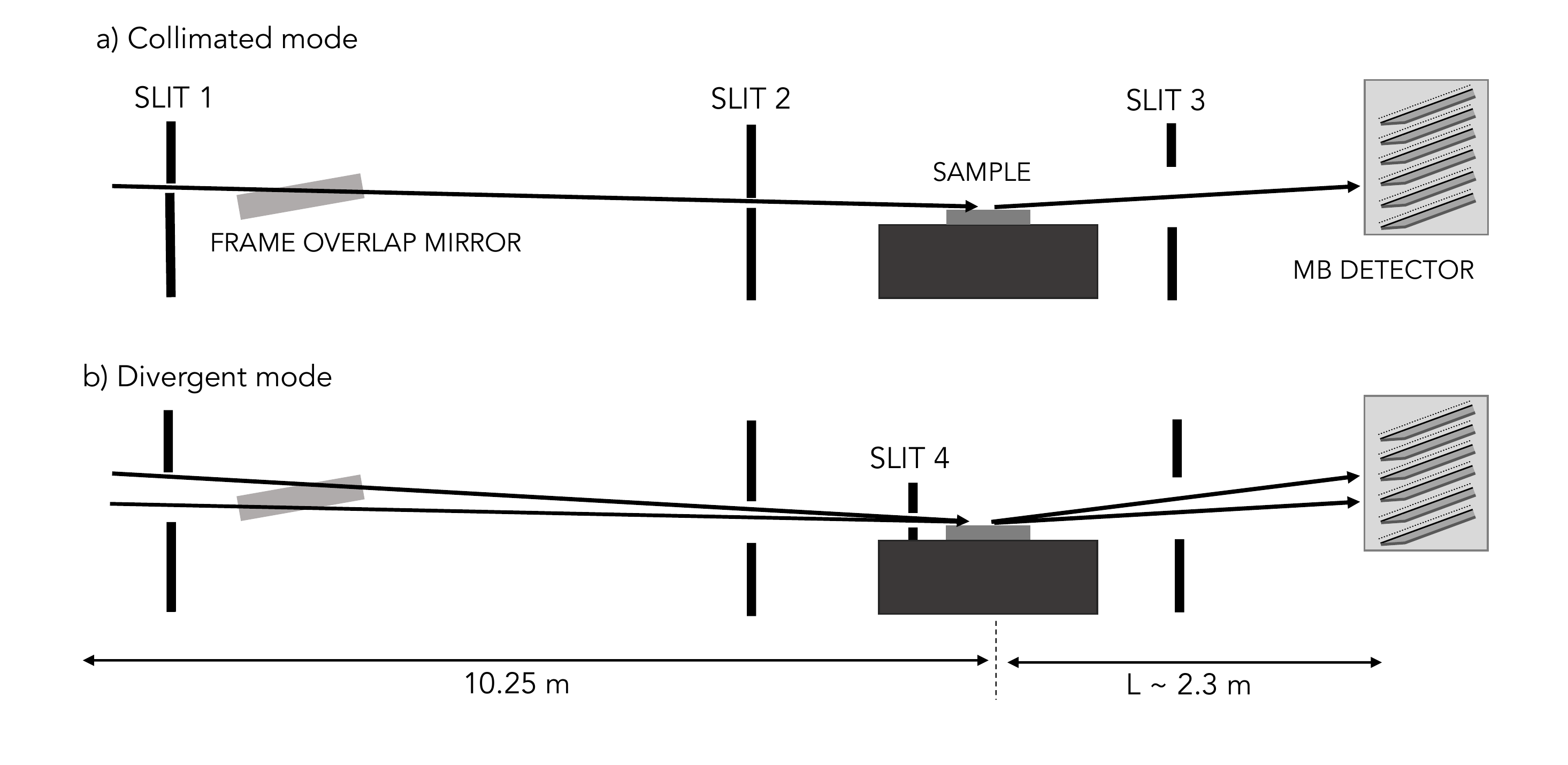}
\caption{\label{fig4bis} \footnotesize A sketch of the CRISP horizontal reflectometer and the MB detector showing the orientation of the cassettes. The beam can be collimated at the sample position either with a low divergence (collimated mode a) or allowing more divergence of the beam (divergent mode b).}
\end{figure} 
\\The beam can be well-collimated using adjustable slits along the beam line, a sketch is shown in figure~\ref{fig4bis}. According to the position and the opening of the slits, we performed the measurements in two working modes: collimated and divergent. In the collimated mode the slits are set in order to achieve a good collimation of the beam at the sample. The divergent mode is obtained opening as much as possible the slits before the sample. According to the concept of REFocus~\cite{OTT_refocus}, proposed for ESTIA~\cite{INSTR_ESTIA0,INSTR_ESTIA1}, one more slit with a narrow opening ($\approx 1$ mm) was added before the sample as shown in figure~\ref{fig4bis}. Figure~\ref{fig46} shows the Multi-Blade installed on CRISP reflectometer.
\begin{figure}[htbp]
\centering
\includegraphics[width=.6\textwidth,keepaspectratio]{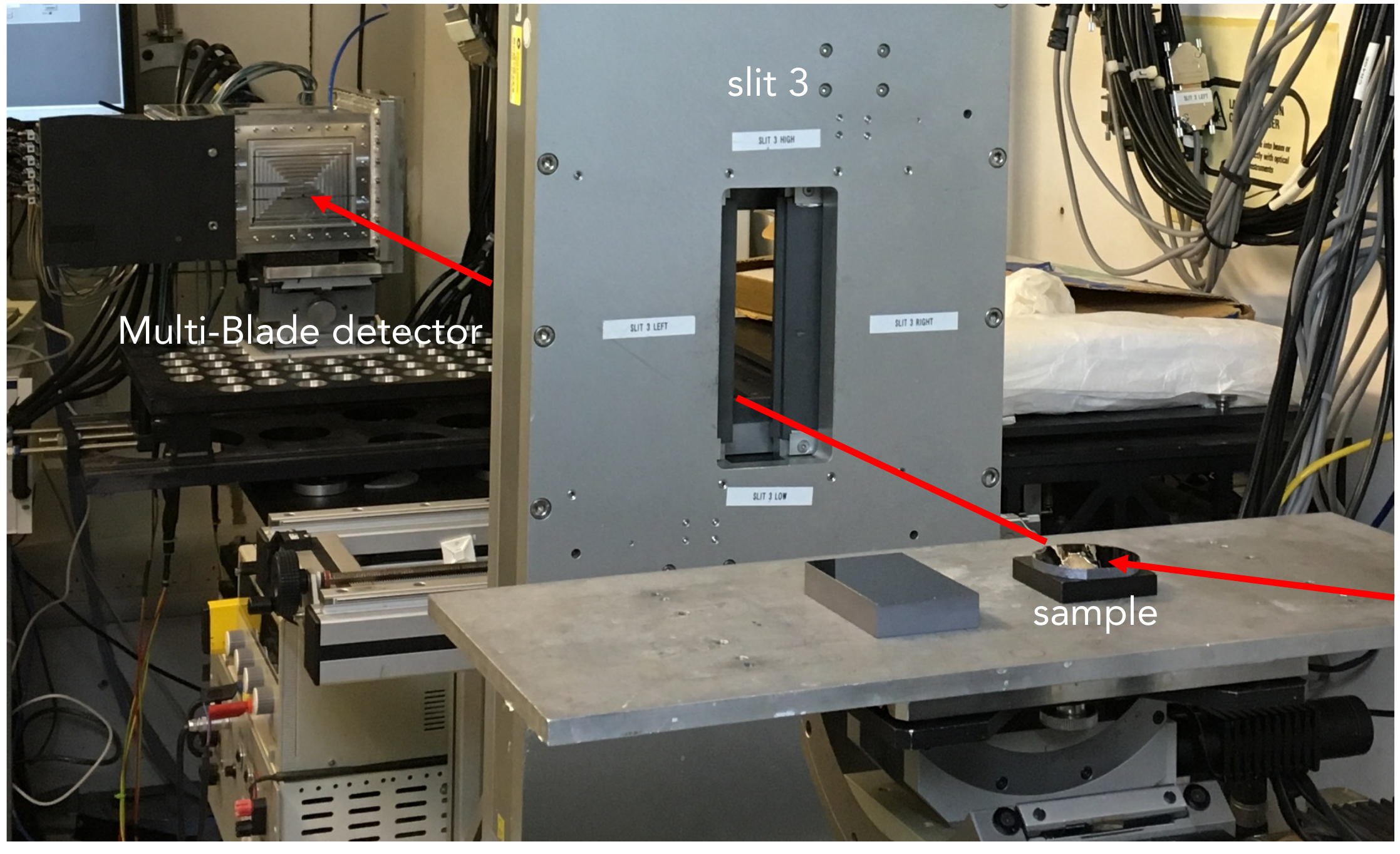}
\caption{\label{fig46} \footnotesize The MB installed on the table of CRISP on a goniometer. A view of the incoming and the reflected beam reaching the active area of the Multi-Blade detector is shown.}
\end{figure} 
\\Three standard and well-known samples have been used in the tests: an iridium (Ir) sample deposited on a silicon substrate (4 $\times$ 4 cm$^2$), a bare silicon (Si) sample ($\approx$ 8 cm diameter) and a Fe/Si super-mirror ($\approx$ 4 cm diameter) which is used in neutron optics to deliver neutrons to the instruments. The Ir sample has been used to study the effect of the spatial resolution of the detector on the measured reflectivity curve and it will be shown in section~\ref{secIr}. The Si sample has been used to study the collimated and divergent modes. This will be discussed in details in section~\ref{colldivsi}. The Fe/Si super-mirror has been used to study the off-specular scattering with the Multi-Blade and it will be discussed in section~\ref{offfesi}.

\section{Results}

\begin{figure}[htbp]
\centering
\includegraphics[width=.32\textwidth,keepaspectratio]{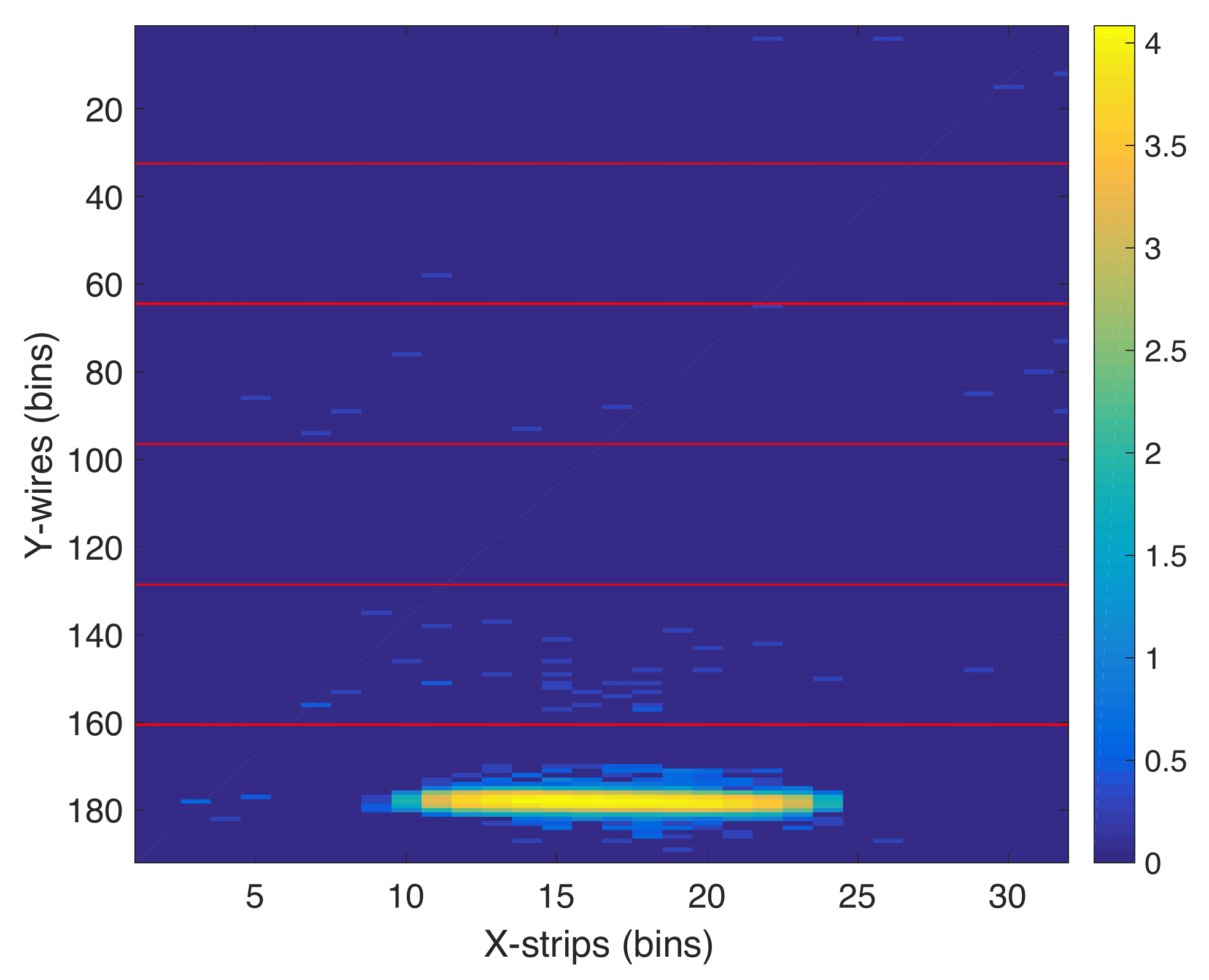}
\includegraphics[width=.32\textwidth,keepaspectratio]{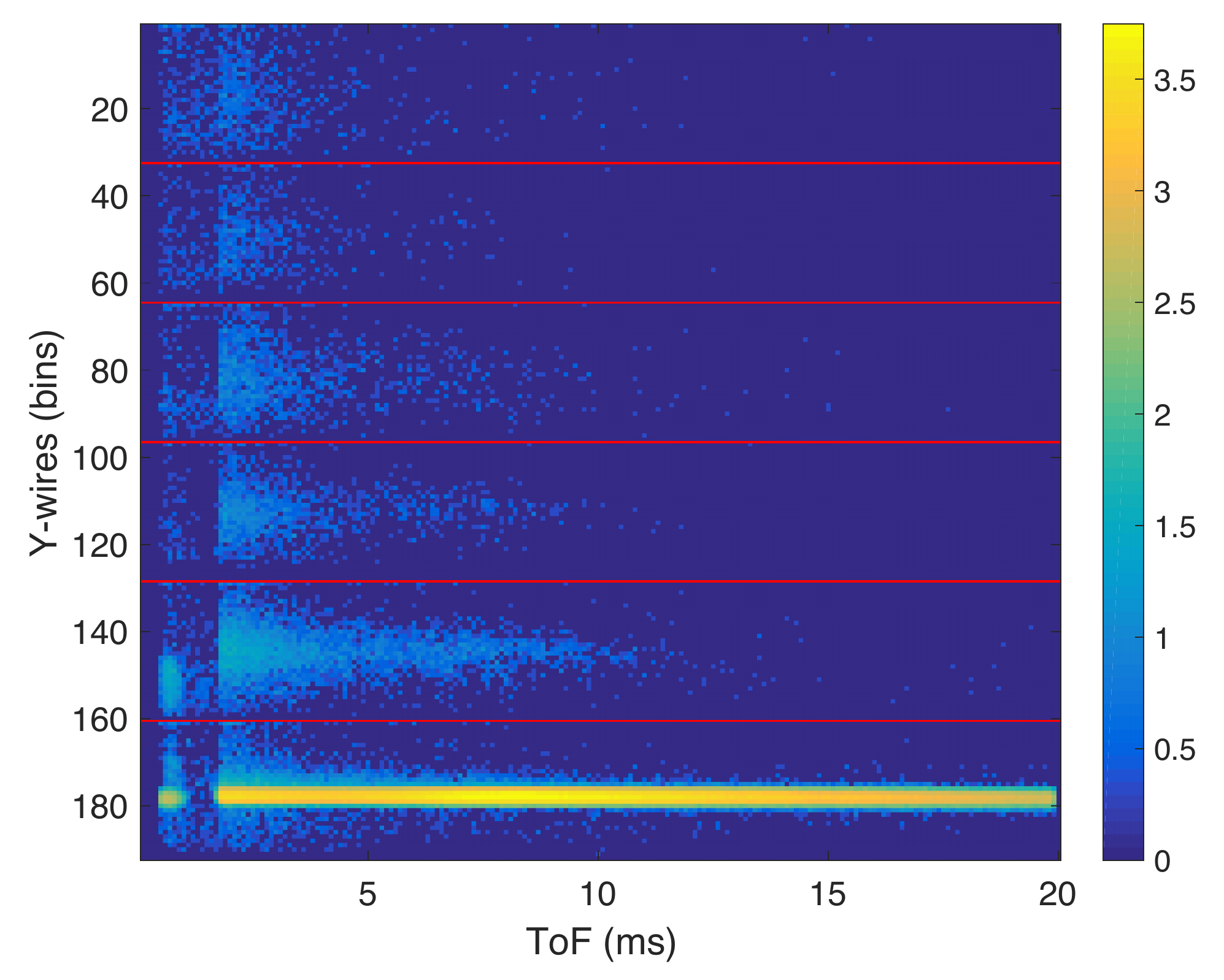}
\includegraphics[width=.32\textwidth,keepaspectratio]{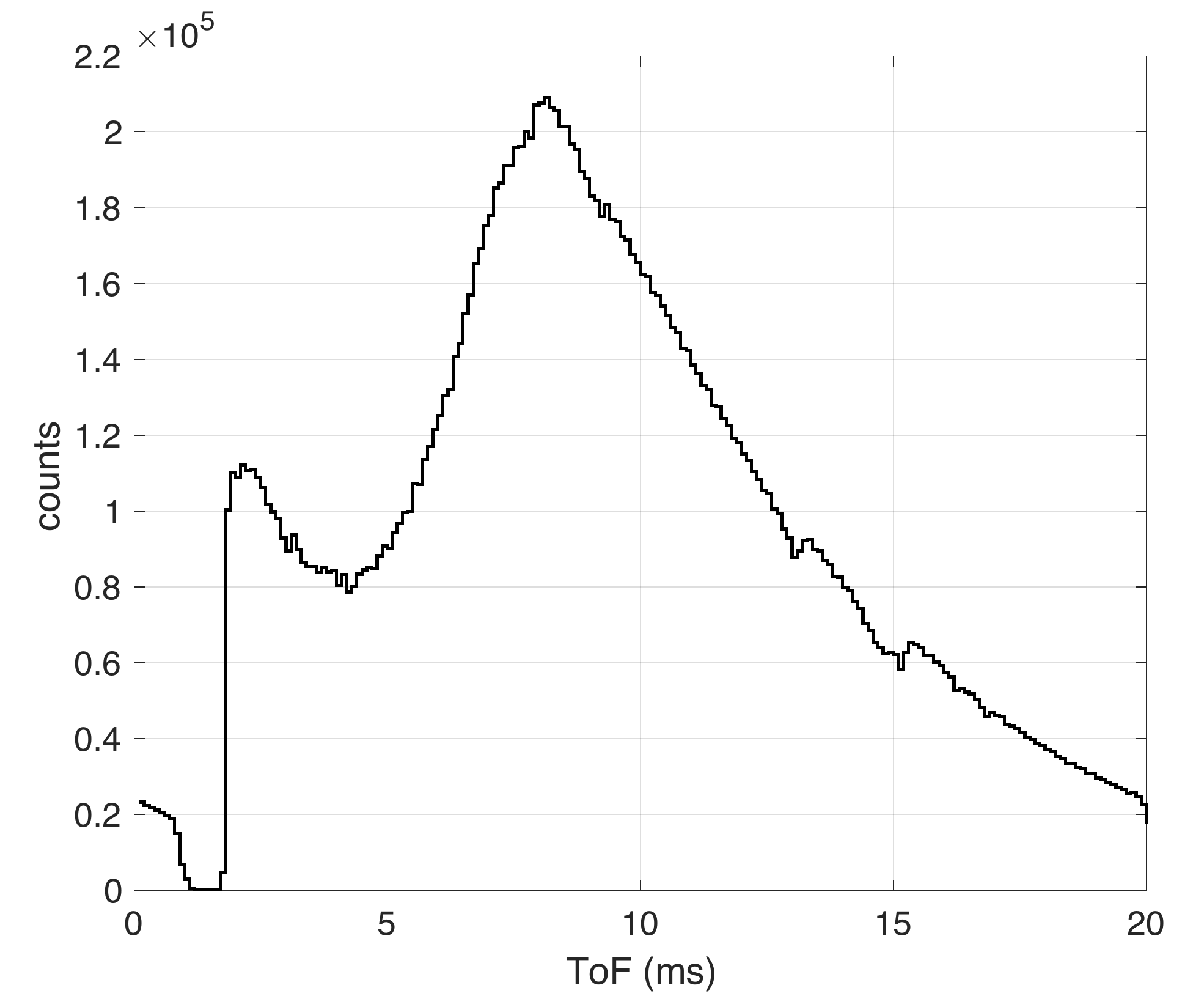}
\caption{\label{fig4} \footnotesize Left: 2D image of the direct beam impinging on the lower cassette of the detector. A gate in ToF, between $12.5\,ms$ and $20\,ms$ (4\AA\,- 6.5\AA), is applied. The bin size on the $Y$ axis is 0.35 mm and on the $X$ axis is 4 mm. Center: ToF image of the detector integrated over the $X$ coordinate. The bin size on the $Y$ axis is 0.35 mm and 100 $\mu$s on the $ToF$ axis. The color bar represents counts in logarithmic scale. Right: intensity of the direct beam in ToF, integrated over the $X$- and gated in the $Y$-coordinate.}
\end{figure}

The triplets $(X,Y,ToF)$ that identify a neutron event, can be represented by two-dimensional plots: the 2D image of the detector is reproduced by the $(X,Y)$ coordinates and the ToF image of the detector which corresponds to the $(Y,ToF)$ coordinates integrating over the other spatial coordinate ($X$). Moreover, the 2D image $(X,Y)$ can be either integrated over the ToF coordinate or gated in any range of time. The ToF image can be integrated or gated over the spatial coordinates as well. A ToF of $6\,$ms corresponds approximately to 1.8\AA, $8\,$ms to 2.5\AA\, and $12.5$ms to 4\AA. An example of these plots is shown in figure~\ref{fig4} and corresponds to a measurement of the direct beam hitting the lower cassette. The 2D image ($(X,Y)$ in logarithmic scale) of the direct beam, gated in ToF between 12.5 ms and 20 ms (4\AA\,- 6.5\AA), is shown in figure~\ref{fig4} on the left. The horizontal red lines indicate where each cassette starts and ends.
The ToF imagine ($(Y,ToF)$ in logarithmic scale) is shown in the center of figure~\ref{fig4}. The ToF integrated over the $X$ coordinate and gated in the $Y$ coordinated around the direct beam area is shown in figure~\ref{fig4} on the right. This is used to normalize the reflectivity measurements of the samples described in the following sections.

The gate in ToF is applied in order to reject the background arising from the spurious scattering from the substrate of the cassettes (the blades). This effect is due to the neutrons that cross the $\mathrm{^{10}B_4 C}$ layer without being absorbed. They are scattered by the substrate and detected in the other cassettes. This background has been understood quantitatively and its full characterization is explained in detail in~\cite{MIO_MB16CRISP_jinst}. Although this effect can be minimized during the analysis by applying the gate in ToF mentioned above, it can be avoided with technical measures that will be implemented in the next detector generation~\cite{MIO_MB16CRISP_jinst}.
As mentioned above the triplets define a three-dimensional space, the third coordinate (Z), which is integrated over, describes the physical position of each wire in depth. Since this position is known, the flight path $D$ can be corrected with the distance ($Z_i$) of the wire $i-th$ of each cassette according this formula:  
\begin{equation}
\label{equadep}
D_i = D_0+Z_i = D_0+(Y_i - 1)\cdot(p\cdot\cos(\beta))
\end{equation}
where $D_0$ depends on the instrument geometry and in our case is the distance from moderator to the first wire (front wire) of the Multi-Blade corresponding to $Y_1=1$, $p=4\,$mm is the wire pitch and $\beta=5^o$ is the inclination of each blade with respect to the sample position.  

\subsection{Specular reflectometry on Ir sample: improvement of the q-resolution with the detector spatial resolution}\label{secIr}

\begin{figure}[htbp]
\centering
\includegraphics[width=.49\textwidth,keepaspectratio]{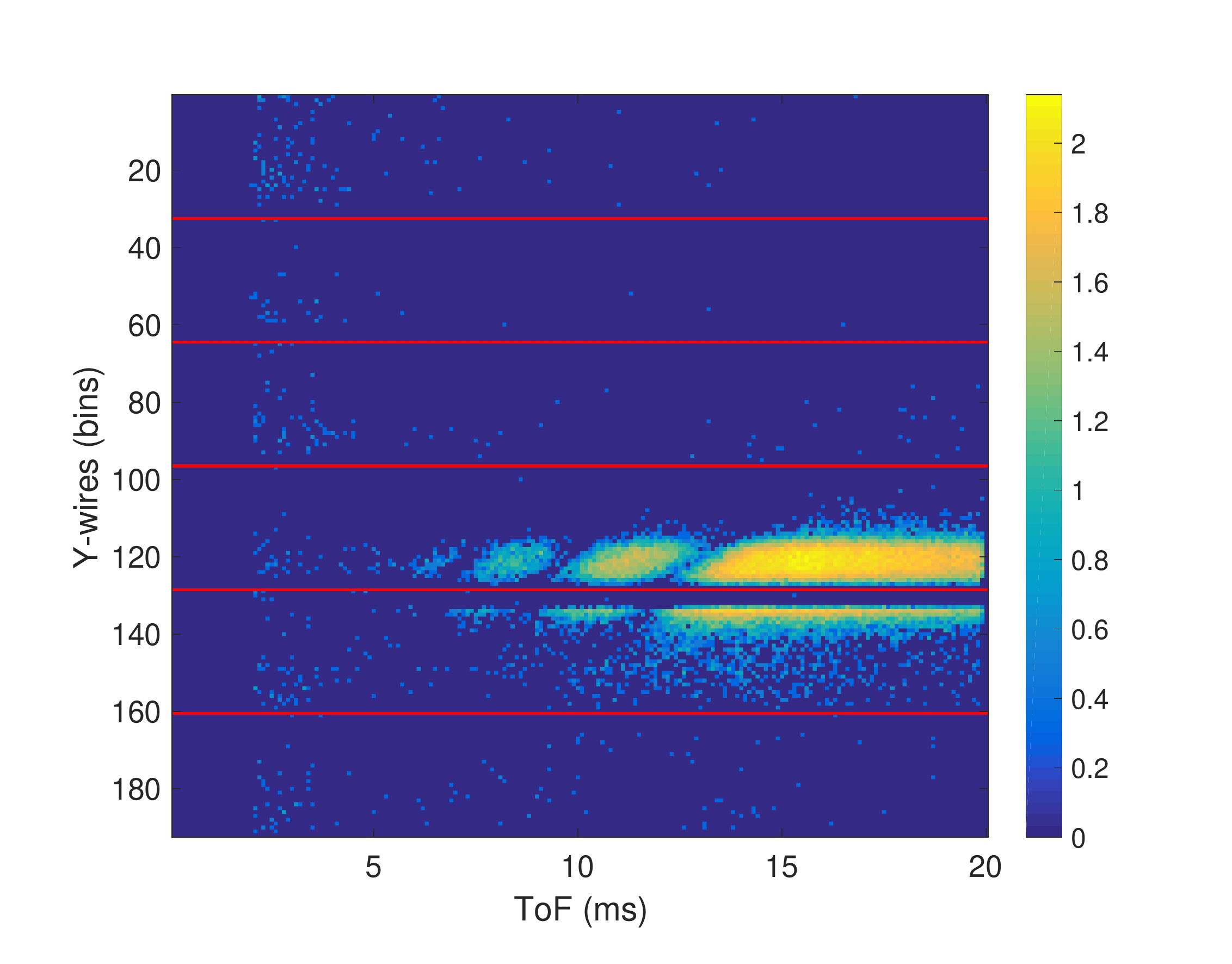}
\includegraphics[width=.49\textwidth,keepaspectratio]{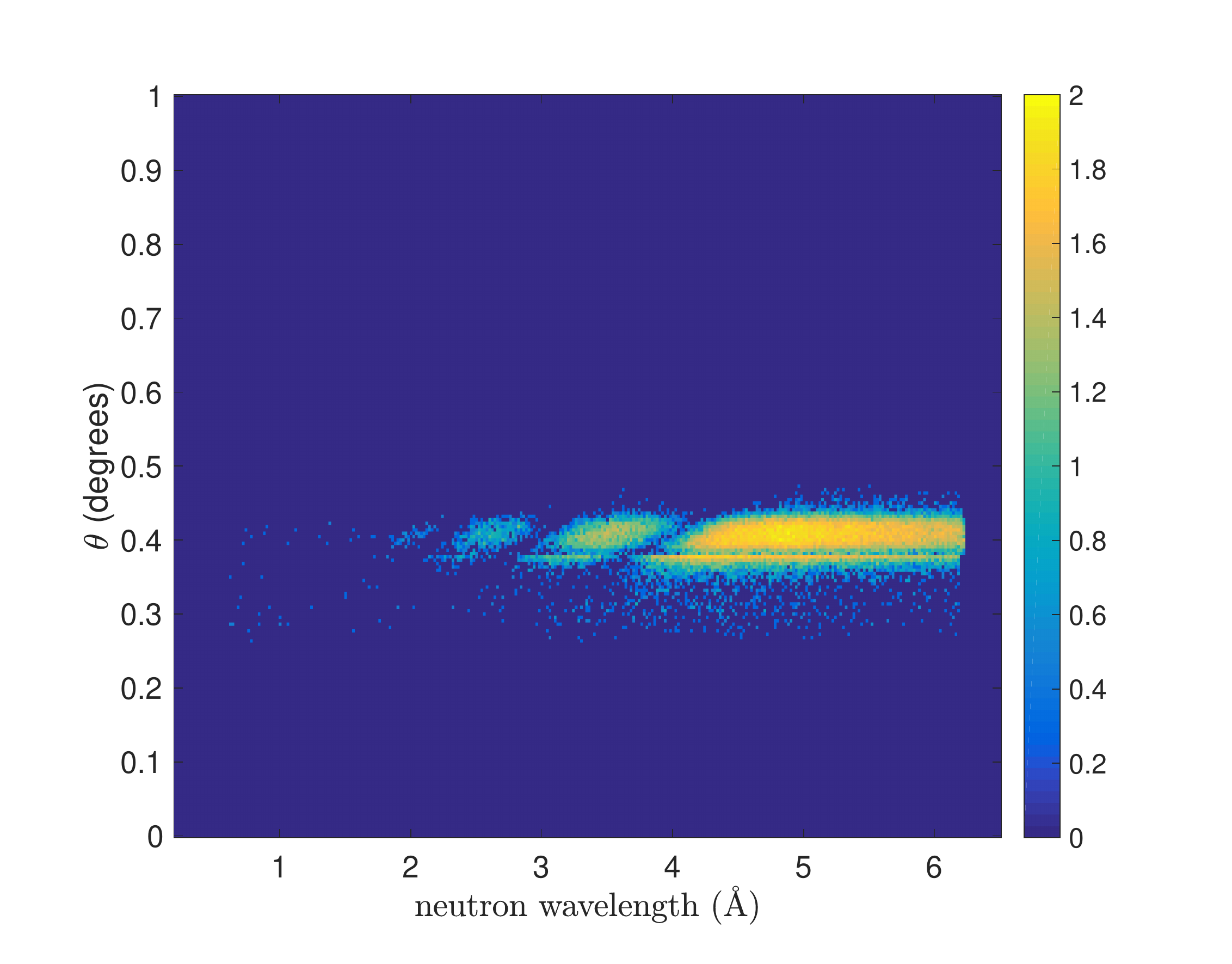}
\caption{\label{lt} \footnotesize Left: ToF spectrum of the reflected beam from the Ir sample. The bin size on the $Y$ axis is 0.35 mm and 100 $\mu$s on the $ToF$ axis. The horizontal lines depict the end of each cassette and the gap in between is the shadowing effect due to geometric properties of the detector. Right: ToF spectrum reduced in the $(\theta, \lambda)$ space. The gap does not represent a dead area, thus can be removed without losing information as shown for the reduced data in $(\theta, \lambda)$ space. The color bar represents counts in logarithmic scale.}
\end{figure}

An iridium (Ir) sample has been used to perform measurements of specular reflectivity. The aim of this measurement was to show how the data analysis can be improved, if the detector spatial resolution is taken into account, and how a finer spatial resolution affects the quality of the results.
\\The reflected intensity from the Ir sample in the $(Y,ToF)$ coordinates is shown on the left graph of figure~\ref{lt}.

The horizontal lines represent the boundaries of each cassette and the gap in between is a shadowing effect caused by the arrangement of the blades. Two subsequent cassettes are arranged in order to have an overlapping area, therefore the gap is not a dead area of the detector. The last firing wire of one cassette, not necessarily the last physical wire, is, in the projected space $(X,Y)$, the neighbour of the first wire of the adjacent cassette. Thus the gap can be removed without losing any information. Moreover, due to the blade geometry the gas gain differs for different wires within a cassette as shown in~\cite{MIO_MB2017, MIO_MB16CRISP_jinst}. The gain drops in the first 7 wires, but it can be compensated by adjusting individual thresholds on each channel. At the very first wire the loss in efficiency corresponds to a drop of $50\%$ with respect to the nominal efficiency. This region of reduced sensitivity is where two cassettes overlap and it is about 0.5 mm wide as shown in figure~\ref{fig99}.
\\On the right of figure~\ref{lt} the $(\theta, \lambda)$ phase space obtained from the $(Y,ToF)$ space is shown. Note that in this plot the gaps have been removed and the sole reduced sensitivity area is still visible in the plot. The neutron wavelength ($\lambda$) is calculated from the ToF corrected with the depth of the detector (equation~\ref{equadep}), thus the exact neutron wavelength can be calculated. 
\\According to equation~\ref{eqaf1}, the wave-vector transfer $q_z$ depends on $\theta$ (determined by the instrumental settings) and $\lambda$. The maximum intensity correspond to the angle between the scattered beam and the sample, $\alpha_f$, being equal to the incident angle $\alpha_f=\alpha_i=\theta$. According to the conventional analysis, for each wavelength, $q_z$ is calculated with a fixed and unique $\theta$ following the equation~\ref{eqaf1} and integrating the intensity over the full size of the beam. The width of the reflected intensity is defined in a range $\alpha_f=\theta \pm \Delta \theta$. The latter originates from the divergence of the beam. 
\\The spatial resolution of the detector can be used to include a correction over $\theta$, as for a small projected sample size this position directly correlates with the reflection angle. This can be used to correct for the increased spread of the reflected beam caused by a slight curvature of the sample surface, which would otherwise reduce the q-resolution.
In contrast to the conventional analysis, each value of $q_z$ is calculated according to its relative $\theta_i = \alpha_i + \delta \theta_i$ defined by the position on the detector. The correction is shown in equation~\ref{thetacorr}:

\begin{equation}
\theta_i = \alpha_i + \delta \theta_i = \alpha_i + f \cdot arctan \Big( \frac{(Y_i-Y_0) \cdot p_s}{L} \Big)
\label{thetacorr}
\end{equation}

where $\mathrm{Y_0} $ is the position of the bin corresponding to $\alpha_f=\alpha_i$, $\mathrm{Y_i}$ is any other position in the integration range, $L$ is the distance between the sample and the detector ($2.3\,$m) and $p_s$ is the pixel size of the detector. Note that the pixel size of the Multi-Blade is $p_s=p\cdot \sin(\beta) \approx 0.34\,$mm, where $p =4\,$mm is the wire pitch, is finer than the spatial resolution of the detector $\approx 0.6\,$mm. The factor $f =1/2$ has to be introduced, as the curvature of the sample surface acts as a change in sample angle and leads to a change in reflection angle by $2\theta$. 
Different combinations of $\lambda$ and $\theta$ correspond to the same $q_z$ in a diagonal cuts of the $(\theta, \lambda)$ space; this leads to an improvement of the resulting reflectivity profile. 
\\ Figure~\ref{lt} clearly visualizes the effect for the bent Ir sample in this manner as it is possible to distinguish three intensity minima from the thickness oscillations that are spread over an extended detector area, much larger than the direct beam. 

\begin{figure}[htbp]
\centering
\includegraphics[width=.7\textwidth,keepaspectratio]{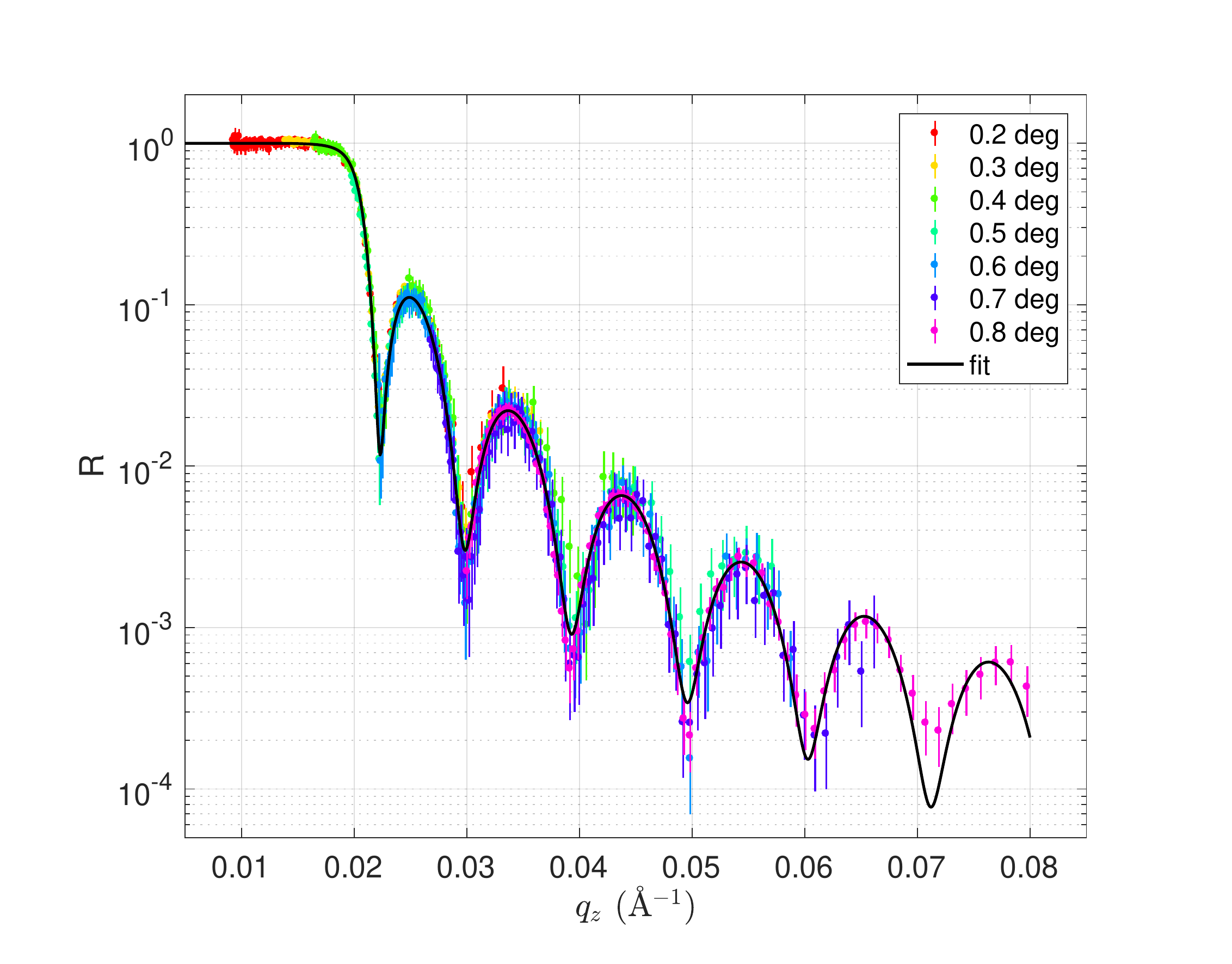}
\includegraphics[width=.7\textwidth,keepaspectratio]{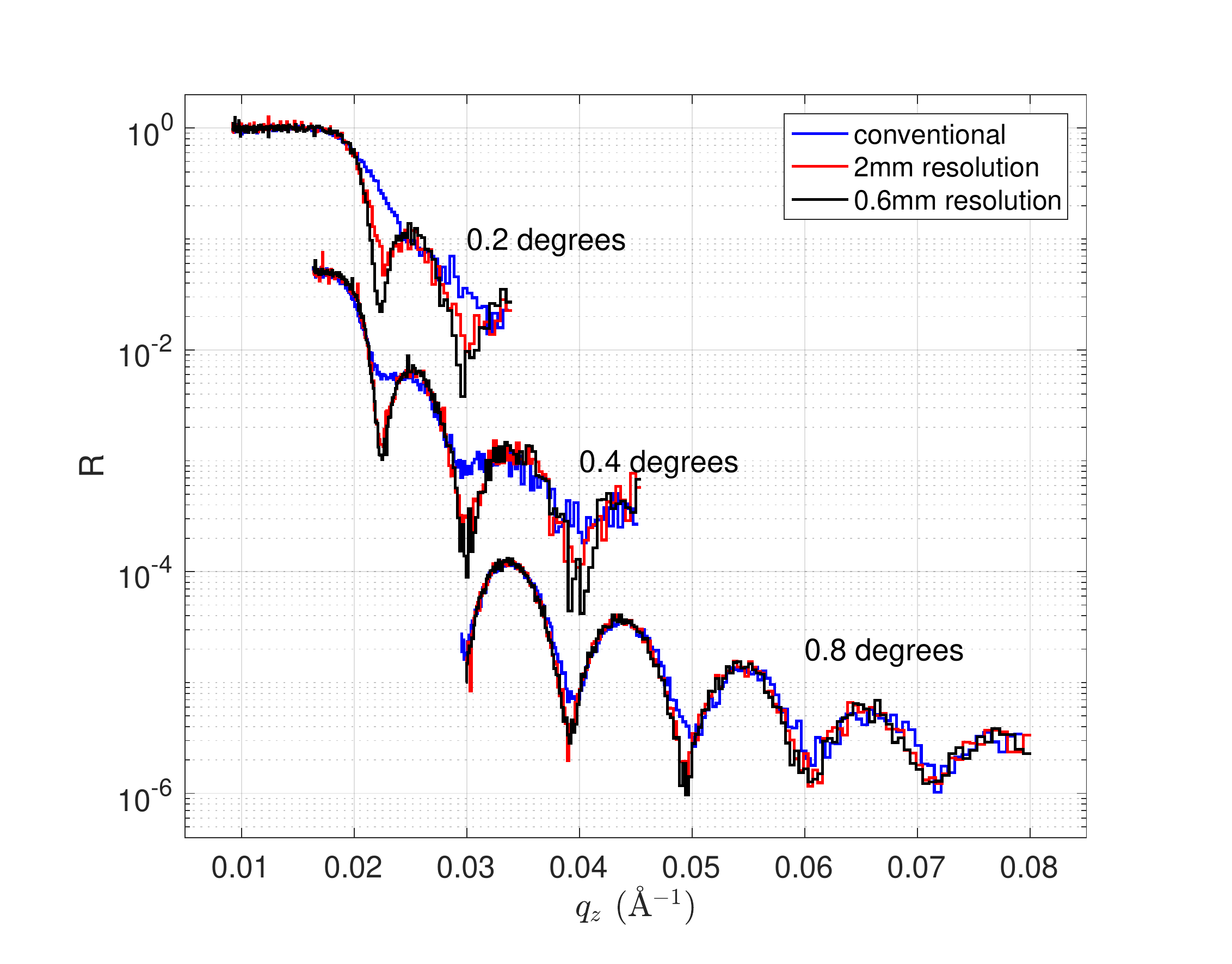}
\caption{\label{figir} \footnotesize Reflectivity curves (R) as a function of the wave-vector transfer ($q$) from an Ir-sample measured with the Multi-Blade detector at several angles and fit (top). Reflectivity curves for the three angles 0.2, 0.4 and 0.8 degrees using the conventional analysis and the $\theta$-corrected analysis with two spatial resolution of the detector, $0.6\,$mm and $2\,$mm (bottom).}
\end{figure}

The sample is a layer of Ir of 550\,\AA{} deposited on a Si substrate. The roughness between the two interfaces is $\approx 10$\AA\ with scattering length density $N_b = 7.3\cdot10^{-6}$\AA$^{-2}$ (see equation~\ref{eqasld}). Figure~\ref{figir} on the top shows the reflectivity curves for several angles used in the measurement, in the range 0.2-0.8 degrees, in steps of 0.1 degrees. The theoretical reflectivity (the fit in the plot) is also shown and it is calculated using the Parratt formalism~\cite{Parratt} and is in good agreement with the experimental data.  
\\ A comparison between the conventional analysis and the $\theta$-corrected reduction is shown in figure~\ref{figir} on the bottom for the three angles: 0.2, 0.4 and 0.8 degrees. The $\theta$-correction was applied considering two pixel sizes, the actual Multi-Blade resolution and a reduced $\approx 2\,$mm resolution which is the current limit of state-of-art detectors used in neutron reflectometry. 
\\At smaller angles, the $q$-resolution depends on the detector spatial resolution to a larger extent. By applying the conventional analysis, the fringes at 0.2 and 0.4 degrees are less visible than if the $\theta$-corrected analysis is used as the better spatial resolution of the detector leads to deeper fringes. 

\subsection{Specular reflectometry on Si sample: dynamic range, collimated and divergent modes}\label{colldivsi}

\begin{figure}[htbp]
\centering
\includegraphics[width=.45\textwidth,keepaspectratio]{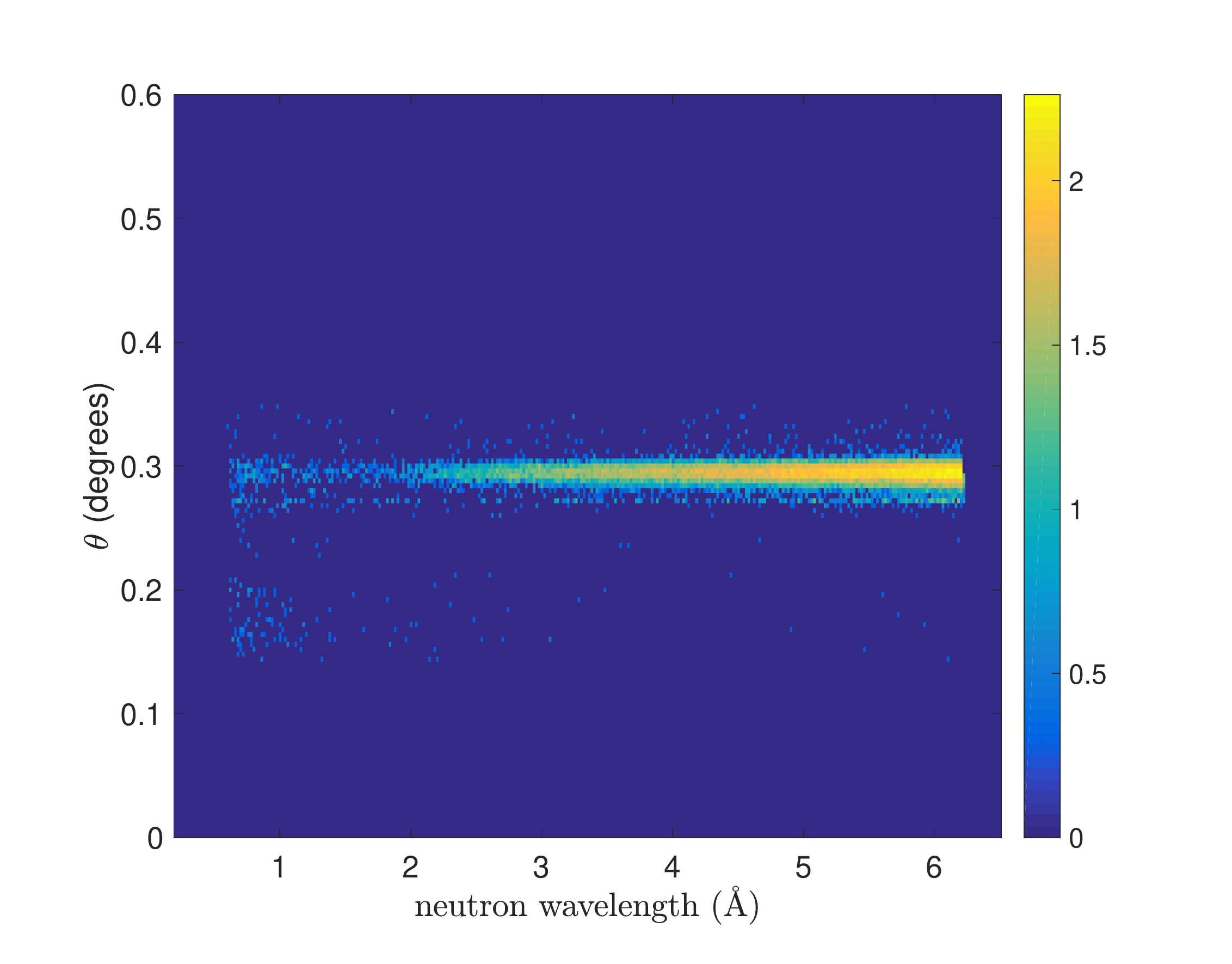}
\includegraphics[width=.45\textwidth,keepaspectratio]{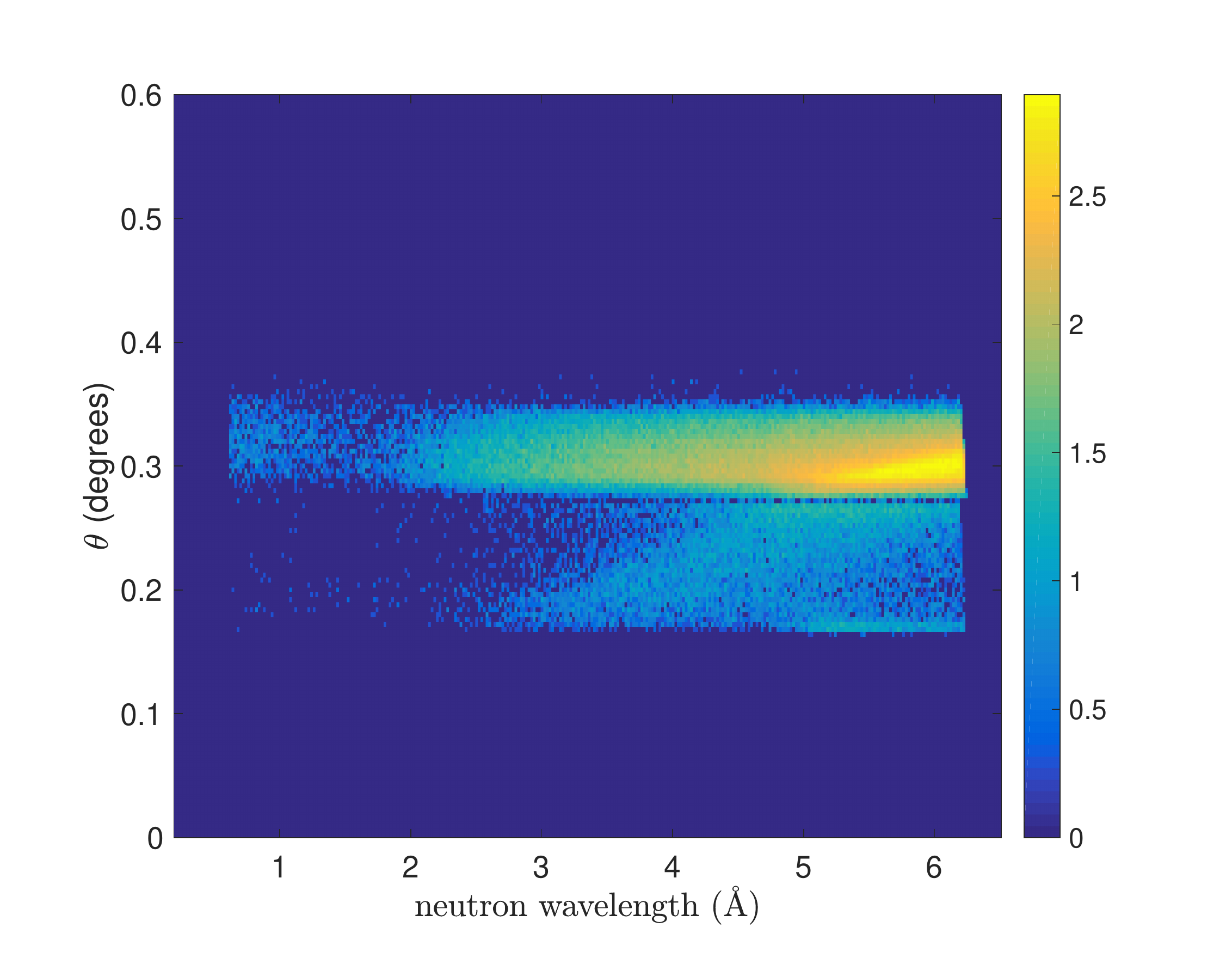}
\caption{\label{figsi1} \footnotesize The $(\theta, \lambda)$ space for the reflectivity of the Si sample at $0.3$ degrees in the collimated (left) and divergent (right) configuration. The color bar represents counts in logarithmic scale.}
\end{figure}

\begin{figure}[htbp]
\centering
\includegraphics[width=.8\textwidth,keepaspectratio]{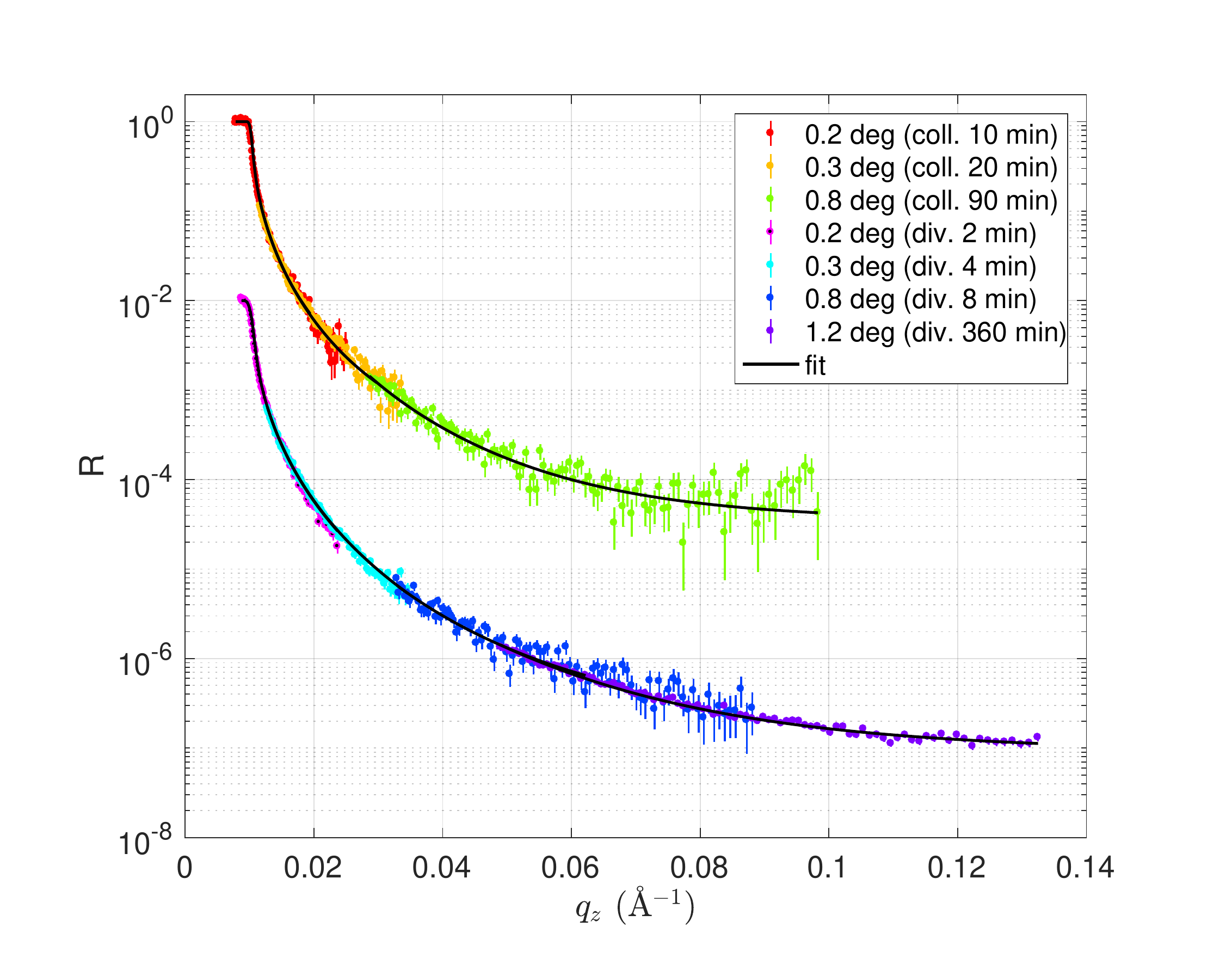}
\caption{\label{figsi2} \footnotesize Specular reflectivity (R) as a function of the wave-vector transfer ($q$) of the Si sample obtained with the collimated and divergent modes. The curves obtained with the divergent mode are shifted by 0.01 in R for clarity.}
\end{figure}

The aim of the measurements presented in this section is to demonstrate the Multi-Blade detector capabilities in a setup as similar as possible to the ESTIA working configurations as described in section~\ref{rifle}. The instrument was operated in two configurations (collimated and divergent modes, see figure~\ref{fig4bis}) and measuring the specular reflectivity from a Si sample.
\\The collimated mode is the conventional working configuration of a reflectometer, where the divergence of the beam is limited due to the slit settings and typically its contribution to the $q$-resolution is set similar to the $\lambda$ contribution.
\\On the other hand the divergent mode exploits the full divergence available at the instrument by only constraining parts of the beam that would not impinge on the sample with the slits. The position of the neutron on the detector is used to encode $\theta$ in a similar manner as described in the previous section, according to equation~\ref{thetacorr}.
%\begin{equation}
%\theta_i = \alpha_i + \delta \theta_i = \alpha_i + \cdot arctan \Big( \frac{(Y_i-Y_0) \cdot SR}{L} \Big)
%\label{thetacorr}
%\end{equation}
Now the factor $f$ is not needed as the sample surface is flat and the change in reflection angle corresponds to the same change in incidence angle. By allowing a wider divergence of the beam, the sampled $\theta$-range is also larger; the available flux at sample increases and thus the measuring time is reduced. This method for the data reduction refers to the one that will be used with ESTIA to allow measurement from very small samples. A detailed description is reported in~\cite{INSTR_ESTIA2}.
\\Although the geometry used for these measurements on CRISP is only an approximate reproduction of the focusing concept used in ESTIA~\cite{INSTR_ESTIA2}, it is useful to test the effectiveness of the Multi-Blade detector response. Note that the focusing obtained with the slits instead of a focusing guide, leads to lower signal and a higher background as the available divergence is smaller and the sample area is strongly over illuminated~\cite{INSTR_ESTIA2}. 
\\ The measurement of specular reflectivity was performed in either configurations on a Si sample at three angles (0.2, 0.3, 0.8 degrees). A further measurement at 1.2 degrees was performed for the divergent mode to reach a wider dynamic range. 
\\ In figure~\ref{figsi1} the intensity of the beam in the $(\theta, \lambda)$ space in the collimated (left) and the divergent (right) modes are shown. The illuminated area of the detector is about 5 times larger in the divergent mode than that of the collimated mode.
\\Figure~\ref{figsi2} depicts the extracted reflectivity of the sample in the two configurations. The total acquisition time for the three angles in the collimated mode is 120 minutes. The same result is obtained in 14 minutes by performing the measurements in the divergent mode. The acquisition time is thus improved by about one order of magnitude.
Despite the high background due to the poor shielding of the Multi-Blade setup on CRISP, a dynamic range of $\approx$ 4 order of magnitude with the three angles was achieved. With a further measurement at 1.2 degrees, we achieved one extra order of magnitude in the dynamic range, which is shown in figure~\ref{figsi2}. Five orders of magnitude is the dynamic range typically reached on this instrument~\cite{INSTR_OSMOND_CRISP}.

It is expected with the Multi-Blade to measure a deeper dynamic range in a better shielded instrument operating environment.

\subsection{Off-specular scattering: Fe/Si supermirror sample}\label{offfesi}
\begin{figure}[htbp]
\centering
\includegraphics[width=.45\textwidth,keepaspectratio]{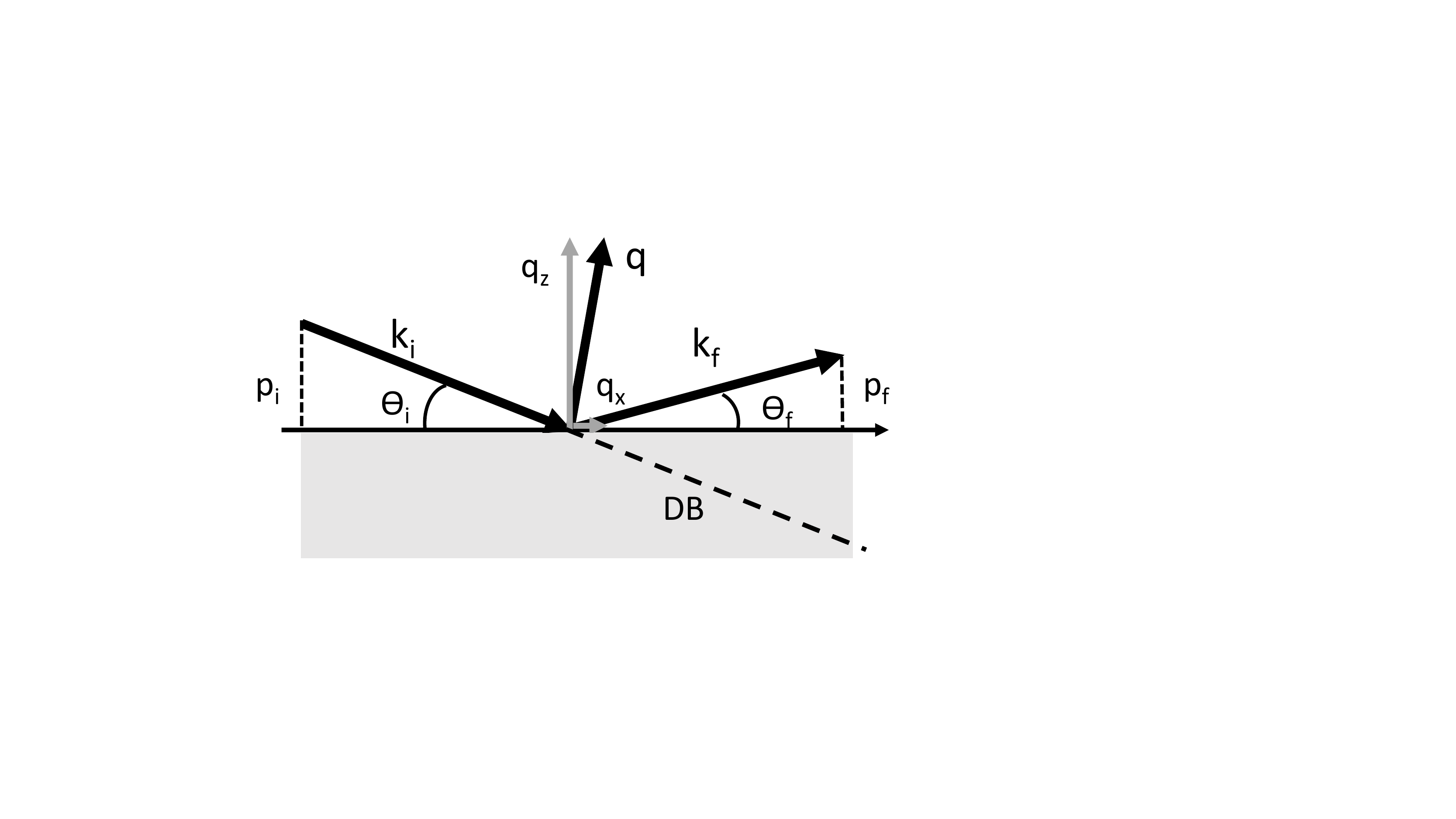}
\caption{\label{fig70} \footnotesize Sketch of the wave vectors definition used in the off-specular scattering.}
\end{figure}

The specular reflectivity allows to probe the structure of a sample across the depth, indeed the scattering vector $q$ is perpendicular to the sample surface. It is possible to probe the in-plane structure of a sample introducing a small parallel component of the scattering vector~\cite{R_offspec0_Ott}; a sketch is shown in figure~\ref{fig70}. The parameter used to reproduce the results of the off-specular scattering are the components of $q$ and the projections of the initial and final wave vectors, they are reported in the equation~\ref{Qcomp}. 

\begin{align}
\label{Qcomp}
&p_i = \frac{2 \pi}{\lambda} \sin \alpha_i & \nonumber \\
&p_f = \frac{2 \pi}{\lambda} \sin \alpha_f & \\
&q_x = \frac{2 \pi}{\lambda} (\cos \alpha_f - \cos \alpha_i)& \nonumber\\
&q_z = \frac{2 \pi}{\lambda} (\sin \alpha_f + \sin \alpha_i)& \nonumber
\end{align}

Neutron off-specular scattering probes the in-plane structure at the $\mu$m length scale. The limitation of this technique is set by both the limited available neutron flux and the small scattering probability. Similarly correlations at the nm length scale can be reached with a collimated beam in both directions, so called grazing incidence small-angle scattering (GISANS), which is described in detail in~\cite{INSTR_GISANS1,INSTR_GISANS2,INSTR_GISANS3,Lauter2016}.
On magnetic samples the off-specular technique allows the depth resolved measurement of correlations from magnetic domains as used in~\cite{Josten2010,Nickel2001}.
\\ Several specific areas can be identified in the off-specular scattering, based on the direction of the final wave vector determined by the reflected angle~\cite{R_offspec1_Ott}. The horizon is defined as $\alpha_f = 0$, when the neutron beam is parallel to the surface of the sample. The specular reflection is found at  $\alpha_i = \alpha_f$ and all other areas above the horizon mark the off-specular scattering region. The direct beam, DB in figure~\ref{fig70}, meets the condition $\alpha_f = - \alpha_i$. When the incident angle is close to the critical angle $\alpha_c$, the transmitted beam is also refracted and thus this equality does not hold for small $\alpha_i$. 
Finally, at $\alpha_i=\alpha_c$ and $\alpha_f=\alpha_c$ one finds the so called Yoneda wings, which are results of dynamic effects mostly produced from surface roughness and magnetic domains.
\begin{figure}[htbp]
\centering
\includegraphics[width=.8\textwidth,keepaspectratio]{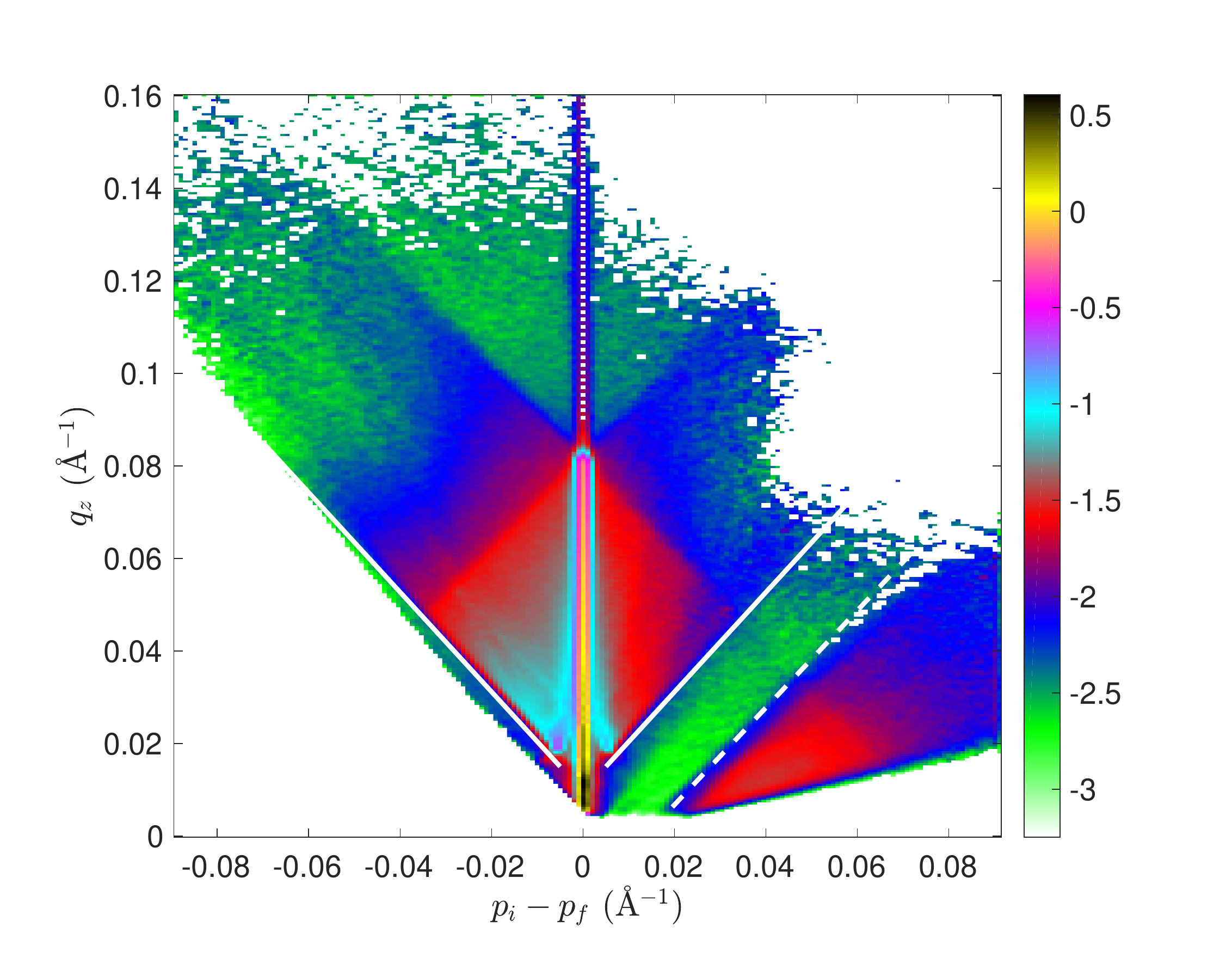}
\caption{\label{figoffsp} \footnotesize Off-specular scattering, expressed as $q_z$ as a function of $p_i - p_f$, from Fe/Si super-mirror: the solid lines corresponds to the two Yoneda wings, the dashed line denotes the beginning of a region of scattered neutrons in the transmission direction (anti-Yoneda), while the dotted line marks the specular reflectivity. The correlated domains from the sample layers correspond to the blue and red rhombus shaped area. }
\end{figure} 
\\ The sample employed to carry out the measurements was a super-mirror Fe/Si (m = 3.8). It shows a strong off-specular scattering when un-magnetized due to complex magnetic domain structures. The off-specular measurements are used to test the detector performances. Good uniformity and spatial resolution as well as large dynamic range are needed to fully characterize the features of the off-specular scattering on the sample.
\\ We performed some measurements using the collimated beam to scan the sample in angle in order to get a fine tuning in uniformity and reach a wider q-space. The sample position was tilted in step of 0.01 degree in the 0.2-0.8 degrees range. The measurements were performed over the whole night. 
The data are presented in the ($p_i - p_f$, $q_z$) coordinates in figure~\ref{figoffsp} and the typical features of the sample are well reproduced. In figure~\ref{figoffsp} the solid lines corresponds to the two Yoneda wings, the dashed line denotes the beginning of a region of scattered neutrons in the transmission direction (anti-Yoneda), while the dotted line marks the specular reflectivity. In the specular reflectivity direction both Silicon and the super-mirror edge are identified, (black spot and the crossing point on the top of the line respectively). The correlated domains from the sample layers correspond to the red and light blue rhombus shaped area. 
This almost featureless area corresponds to magnetic spin-flip scattering within the super-mirror that can be separated by polarization-analysis into two asymmetric components as is demonstrated in~\cite{KLAUSER2016}. 
Neither beam polarization nor magnetic field has been employed, therefore the magnetic scattering of all spin-states are summed together, resulting in the rhombus area. Nevertheless the test demonstrate the establishment of the Multi-Blade detector technology for neutron reflectometry application. 

\section{Conclusions and outlook}

The neutron reflectometry technique represents a challenge in terms of instrument design and detection performances. Nowadays several methods have been proposed to increase the incoming flux leading to improvements for specular neutron reflectivity measurements. Along with the instruments operation the detectors response must be refined. The current detector technology is limited mainly as regards the spatial resolution and counting rate capability. The Multi-Blade detector has been proposed as a valid alternative to replace the state-of-art detectors, because of the better performances on both spatial resolution and counting rate capability. The requirements for this technology are set for the ESS reflectometers (ESTIA~\cite{INSTR_ESTIA,INSTR_ESTIA1,INSTR_ESTIA2} and FREIA~\cite{INSTR_FREIA,INSTR_FREIA2}).  Hence, apart from the ESS reflectometers, reflectometers at other facilities can take advantage of employing the Multi-Blade detector technology.
\\ A campaign of scientific measurements has been performed on the CRISP~\cite{CRISP1} reflectometer at ISIS (Science \& Technology Facilities Council in UK~\cite{ISIS}). The reflectivity of several reference sample have been measured operating the instrument in various configurations to reproduce the setup that will be used at the ESS reflectometers. The measurements, not only provide a validation of the Multi-Blade as a mature technology for neutron reflectometry experiments, but it has been shown that the instrument operation was improved using the Multi-Blade as well.
\\The spatial resolution of a detector is, indeed, deeply connected to the achievable $q$-resolution of the instrument. The calculated $q_z$ is a combination of the neutron wavelength and the scattering angle, the latter can be corrected taking into account the spatial resolution of the detector and thus a higher $q$-resolution is achieved. When measuring the specular reflectivity from a sample which shows interference fringes in $q_z$, such as the Iridium on Silicon, the fringes get more visible as the spatial resolution of the detector improves. The result has been compared to a conventional non-position sensitive detector and with a state-of-the-art detector with $2\,$mm resolution. 

It has been shown that the CRISP instrument can be operated, thanks to the Multi-Blade, in the REFocus mode~\cite{OTT_refocus} (divergent mode) which is one of the standard mode foreseen for the ESTIA reflectometer~\cite{INSTR_ESTIA2}. In this configuration, the correction of the scattering angle for calculating $q_z$ is mandatory and the spatial resolution and the counting rate capability of the detector is a key feature. Moreover, from the measurements of the Silicon sample, the $q$-range was measured to five orders of magnitude, reaching the limits of the instrument, despite the high background at the CRISP instrument and the poor shielding of the Multi-Blade detector. 

An off-specular scattering measurement was also performed on a super-mirror Fe/Si multi-layer sample. Neither beam polarization nor magnetic field has been used in order to have a strong off-specular scattering from the sample. The ability of the Multi-Blade to measure, not only specular, but also off-specular scattering was shown. 

The results presented here show that the Multi-Blade detector technology is mature, and ready for implementation on neutron reflectometers.

\enlargethispage{20pt}

\ethics{This research poses no ethical considerations.}

\dataccess{The raw reflectometry data (DOI: 10.6084/m9.figshare.6138986.v1) of the experiment described in this manuscript can be downloaded from \url{https://figshare.com/s/2712231a48e19818c0c9}.}

\aucontribute{ GM and FP have drafted the manuscript. AG, FP, GM and FM have conceived the experiment. GM, FP and AG have analyzed the data. GM, FP, FM, AG, DR, TI and TA have set the experiment on CRISP and collected the data. MA and ILH have designed the mechanics. CH, LR, SS and PS have provided the Boron-10 coatings for the detector. FM, PP and DV have designed the front-end electronics. GM, FP, FM and TI have designed and programmed the data acquisition system. FP and RHW have conceived and initiated the design of the detector for neutron reflectometers at ESS. All authors gave final approval for publication.}

\competing{We have no competing interests.}

\funding{This work is being supported by the BrightnESS project, Work Package (WP) 4.2 (EU Horizon 2020, INFRADEV-3-2015, 676548) and carried out as a part of the collaboration between the European Spallation Source (ESS - Sweden), the Lund University (LU - Sweden), the Link\"{o}ping University (LiU - Sweden) and the Wigner Research Centre for Physics (Hungary) and the University of Perugia (Italy). 
\\ The work was supported by the Momentum Programme of the Hungarian Academy of Sciences under grant no. LP2013-60.
\\ The work originally started in the context of the collaboration between the Institut Laue-Langevin (ILL - France), the Link\"{o}ping University (LiU - Sweden) and the European Spallation Source (ESS - Sweden) within the context of the International Collaboration on the development of Neutron Detectors (www.icnd.org).}

\ack{The authors would like to thank the ISIS detector group for the support during the tests. The authors thank the CRISP instrument scientists R. Dalgliesh and C. Kinane for providing the beam time and the instrument support necessary for this detector test.}

%\disclaimer{Insert disclaimer text here.}

%\bibliographystyle{ieeetr}
\bibliographystyle{RS}
\bibliography{BIBLIODB}

\begin{thebibliography}{99}

\bibitem{MIO_MB2017}
Piscitelli F, Messi F, Anastasopoulos M, Bry{\'s} T, Chicken F, Dian E, Fuzi J,
  H{\"o}glund C, Kiss G, Orban J, Pazmandi P, Robinson L, Rosta L, Schmidt S,
  Varga D, Zsiros T, Hall-Wilton R. 2017  {The Multi-Blade Boron-10-based
  neutron detector for high intensity neutron reflectometry at ESS}. {\em
  Journal of Instrumentation} \textbf{12}, P03013.

\bibitem{MIO_MB2014}
Piscitelli F, Buffet JC, Clergeau JF, Cuccaro S, Gu{\'e}rard B, Khaplanov A,
  Manna QL, Rigal JM, Esch PV. 2014  {Study of a high spatial resolution 10 B
  -based thermal neutron detector for application in neutron reflectometry: the
  Multi-Blade prototype}. {\em Journal of Instrumentation} \textbf{9}, P03007.

\bibitem{MIO_MBproc}
Piscitelli F, Buffet JC, Correa J, Esch PV, Gu{\'e}rard B, Khaplanov A. 2012
  {Study of a 10B-based Multi-Blade detector for neutron scattering science}.
  In Sci. ITN, editor, {\em Conference record of Nuclear Science Symposium and
  Medical Imaging Conference (NSS/MIC) Anaheim} pp. 171--175.

\bibitem{MIO_MyThesis}
Piscitelli F. 2014 - arXiv:1406.3133 {\em Boron-10 layers, Neutron
  Reflectometry and Thermal Neutron Gaseous Detectors}.
PhD thesis Institut Laue-Langevin and University of Perugia.

\bibitem{MIO_MB16CRISP_jinst}
Piscitelli F, Mauri G, Messi F, Anastasopoulos M, Arnold T, Glavic A,
  H{\"o}glund C, Ilves T, Higuera IL, Pazmandi P, Raspino D, Robinson L,
  Schmidt S, Svensson P, Varga D, Hall-Wilton R. 2018  {Characterization of the
  Multi-Blade 10B-based detector at the CRISP reflectometer at ISIS for neutron
  reflectometry at ESS}. arXiv:1803.09589 (submitted to J. Instr.).

\bibitem{INSTR_D17}
Cubitt R, Fragneto G. 2002  D17: the new reflectometer at the ILL. {\em Applied
  Physics A} \textbf{74}, s329--s331.

\bibitem{INSTR_FIGARO}
Campbell RA, Wacklin HP, Sutton I, Cubitt R, Fragneto G. 2011  FIGARO: The new
  horizontal neutron reflectometer at the ILL. {\em The European Physical
  Journal Plus} \textbf{126}, 1--22.

\bibitem{INSTR_ISIS_R}
Charlton TR, Coleman RLS, Dalgliesh RM, Kinane CJ, Neylon C, Langridge S, Plomp
  J, Webb NGJ, Webster JRP. 2011  Advances in Neutron Reflectometry at ISIS.
  {\em Neutron News} \textbf{22}, 15--18.

\bibitem{ESS}

European Spallation Source ESS ERIC - http://europeanspallationsource.se.

\bibitem{INSTR_FREIA}
Wacklin H. 2014 .
FREIA: Reflectometer concept for fast kinetics at ESS - ESS instrument proposal
  -
  {https://europeanspallationsource.se/sites/default/files/freia\_proposal.pdf}.

\bibitem{INSTR_FREIA2}
Wacklin H. 2016 .
Revealing Change Over Time, FREIA Brings Fast Kinetic Studies to Reflectometry
  -
  http://neutronsources.org/news/scientific-highlights/revealing-change-over-time-freia-brings-fast-kinetic-studies-to-reflectometry.html.

\bibitem{INSTR_ESTIA}
Stahn J. 2014  ESTIA: A Truly Focusing Reflectometer. ESS instrument proposal.

\bibitem{INSTR_ESTIA1}
{Stahn, J.}, {Filges, U.}, {Panzner, T.}. 2012  Focusing specular neutron
  reflectometry for small samples. {\em Eur. Phys. J. Appl. Phys.} \textbf{58},
  11001.

\bibitem{INSTR_ESTIA2}
Stahn J, Glavic A. 2016  {Focusing neutron reflectometry: Implementation and
  experience on the TOF-reflectometer Amor}. {\em Nuclear Instruments and
  Methods in Physics Research Section A: Accelerators, Spectrometers, Detectors
  and Associated Equipment} \textbf{821}, 44 -- 54.

\bibitem{MISC_pike2002scattering}
Pike E, Sabatier P. 2002 {\em Scattering: Scattering and Inverse Scattering in
  Pure and Applied Science}.
Number v. 1 in Scattering: Scattering and Inverse Scattering in Pure and
  Applied Science. Academic Press.

\bibitem{R_fermi}
Fermi E, Zinn W, Laboratory LAN, Commission UAE. 1946 {\em Reflection of
  neutrons on mirrors}.
Manhattan District.

\bibitem{Lauter2016}
Lauter V, Lauter H, Glavic A, Toperverg B. 2016  Reflectivity, Off-Specular
  Scattering, and {GISANS} Neutrons. In {\em Reference Module in Materials
  Science and Materials Engineering} pp.~--. Elsevier.

\bibitem{R_offspec0_Ott}
Ott F. 2009 {\em {Neutron scattering on magnetic nanostructures}}.
Condensed Matter [cond-mat]. Universit{\'e} Paris Sud - Paris XI.

\bibitem{OTT_general}
Ott F, Menelle A. 2009  New designs for high intensity specular neutron
  reflectometers. {\em The European Physical Journal Special Topics}
  \textbf{167}, 93--99.

\bibitem{INSTR_R_Spin2}
Major J, Dosch H, Felcher G, Habicht K, Keller T, te~Velthuis S, Vorobiev A,
  Wahl M. 2003  Combining of neutron spin echo and reflectivity: a new
  technique for probing surface and interface order. {\em Physica B: Condensed
  Matter} \textbf{336}, 8 -- 15.
Proceedings of the Seventh International Conference on Surface X-ray and
  Neutron Scattering.

\bibitem{OTT_tiltof}
Ott F, Menelle A. 2006  TilToF: A high-intensity space--time reflectometer.
  {\em Physica B: Condensed Matter} \textbf{385-386}, 985 -- 988.

\bibitem{OTT_gradtof}
Ott F, de~Vismes A. 2007  RefloGrad/GradTOF: Neutron energy analysis for a very
  high-flux neutron reflectometer. {\em Physica B: Condensed Matter}
  \textbf{397}, 153 -- 155.

\bibitem{OTT_refocus}
Ott F, Menelle A. 2008  REFocus: A new concept for a very high flux neutron
  reflectometer. {\em Nuclear Instruments and Methods in Physics Research
  Section A: Accelerators, Spectrometers, Detectors and Associated Equipment}
  \textbf{586}, 23 -- 30.
Proceedings of the European Workshop on Neutron Optics.

\bibitem{R_Cubitt1}
Cubitt R, Stahn J. 2011  Neutron reflectometry by refractive encoding. {\em The
  European Physical Journal Plus} \textbf{126}, 1--5.

\bibitem{R_Cubitt2}
Cubitt R, Shimizu H, Ikeda K, Torikai N. 2006  Refraction as a means of
  encoding wavelength for neutron reflectometry. {\em Nuclear Instruments and
  Methods in Physics Research Section A: Accelerators, Spectrometers, Detectors
  and Associated Equipment} \textbf{558}, 547 -- 550.

\bibitem{INSTR_R_Spin}
Rekveldt MT. 1997  Neutron reflectometry and SANS by neutron spin echo. {\em
  Physica B: Condensed Matter} \textbf{234-236}, 1135 -- 1137.
Proceedings of the First European Conference on Neutron Scattering.

\bibitem{PSI}

{Paul Scherrer Institute (PSI) - https://www.psi.ch }.

\bibitem{INSTR_ESTIA0}
Stahn J, Panzner T, Filges U, Marcelot C, B{\"o}ni P. 2011  Study on a
  focusing, low-background neutron delivery system. {\em Nuclear Instruments
  and Methods in Physics Research Section A: Accelerators, Spectrometers,
  Detectors and Associated Equipment} \textbf{634}, S12 -- S16.
Proceedings of the International Workshop on Neutron Optics NOP2010.

\bibitem{ESS_TDR}
Peggs S.
ESS Technical Design Report - (ESS-2013-0001) -
  {http://eval.esss.lu.se/cgi-bin/public/DocDB/ShowDocument?docid=274}.

\bibitem{DET_rates}
Stefanescu I, Christensen M, Fenske J, Hall-Wilton R, Henry P, Kirstein O,
  M{\"u}ller M, Nowak G, Pooley D, Raspino D, Rhodes N, {\v S}aroun J, Schefer
  J, Schooneveld E, Sykora J, Schweika W. 2017  Neutron detectors for the {ESS}
  diffractometers. {\em Journal of Instrumentation} \textbf{12}, P01019.

\bibitem{HE3S_kirstein}
Kirstein O, et~al.. 2014  Neutron Position Sensitive Detectors for the {ESS}.
  {\em arXiv:1411.6194} \textbf{Proceedings of the 23rd International Workshop
  on Vertex Detectors, 15-19 September 2014, Macha Lake, The Czech Republic}.

\bibitem{CRISP1}

CRISP instrument manual 2010 -
  https://www.isis.stfc.ac.uk/Pages/crisp-instrument-manual-nov-2010.pdf.

\bibitem{ISIS}

ISIS Neutron and Muon Source - https://www.isis.stfc.ac.uk.

\bibitem{B4C_carina}
H{\"o}glund C, Birch J, Andersen K, Bigault T, Buffet JC, Correa J, van Esch P,
  Guerard B, Hall-Wilton R, Jensen J, Khaplanov A, Piscitelli F, Vettier C,
  Vollenberg W, Hultman L. 2012  B4C thin films for neutron detection. {\em
  Journal of Applied Physics} \textbf{111}.

\bibitem{B4C_carina3}
H{\"o}glund C, Zeitelhack K, Kudejova P, Jensen J, Greczynski G, Lu J, Hultman
  L, Birch J, Hall-Wilton R. 2015  Stability of {10B4C} thin films under
  neutron radiation. {\em Radiation Physics and Chemistry} \textbf{113}, 14 --
  19.

\bibitem{B4C_Schmidt}
Schmidt S, H{\"o}glund C, Jensen J, Hultman L, Birch J, Hall-Wilton R. 2016
  Low-temperature growth of boron carbide coatings by direct current magnetron
  sputtering and high-power impulse magnetron sputtering. {\em Journal of
  Materials Science} \textbf{51}, 10418--10428.

\bibitem{EL_CAEN}

CAEN - Electronic Instrumentation - http://www.caen.it.

\bibitem{Parratt}
Parratt LG. 1954  Surface Studies of Solids by Total Reflection of X-Rays. {\em
  Phys. Rev.} \textbf{95}, 359--369.

\bibitem{INSTR_OSMOND_CRISP}
Bateman J, Dalgliesh R, Duxbury D, Helsby W, Holt S, Kinane C, Marsh A, Rhodes
  N, Schooneveld E, Spill E, Stephenson R. 2013  The OSMOND detector. {\em
  Nuclear Instruments and Methods in Physics Research Section A: Accelerators,
  Spectrometers, Detectors and Associated Equipment} \textbf{698}, 168 -- 176.

\bibitem{INSTR_GISANS1}
Kentzinger E, Frielinghaus H, R{\"u}cker U, Ioffe A, Richter D, Br{\"u}ckel T.
  2007  Probing lateral magnetic nanostructures by polarized GISANS. {\em
  Physica B: Condensed Matter} \textbf{397}, 43 -- 46.

\bibitem{INSTR_GISANS2}
Fermon C, Ott F, Gilles B, Marty A, Menelle A, Samson Y, Legoff G, Francinet G.
  1999  Towards a 3D magnetometry by neutron reflectometry. {\em Physica B:
  Condensed Matter} \textbf{267-268}, 162 -- 167.

\bibitem{INSTR_GISANS3}
Pannetier M, Ott F, Fermon C, Samson Y. 2003  Surface diffraction on magnetic
  nanostructures in thin films using grazing incidence SANS. {\em Physica B:
  Condensed Matter} \textbf{335}, 54 -- 58.
Proceedings of the Fourth International Workshop on Polarised Neutrons for
  Condensed Matter Investigations.

\bibitem{Josten2010}
Josten E, Rucker U, Mattauch S, Korolkov D, Glavic A, Bruckel T. 2010
  Magnetization Flop in {Fe/Cr} {GMR} Multilayers. {\em Journal of Physics
  Conference Series} \textbf{211}, UNSP 012023.
Polarized Neutrons and Synchrotron X-rays For Magnetism Conference 2009.

\bibitem{Nickel2001}
Nickel B, R{\"u}hm A, Donner W, Major J, Dosch H, Schreyer A, Zabel H, Humblot
  H. 2001  Spin-resolved off-specular neutron scattering maps from magnetic
  multilayers using a polarized 3He gas spin filter. {\em Review of Scientific
  Instruments} \textbf{72}, 163--172.

\bibitem{R_offspec1_Ott}
Kozhevnikov SV, Ott F. 2010  Representation of data on off-specular neutron
  scattering. {\em Physics of the Solid State} \textbf{52}, 1561--1570.

\bibitem{KLAUSER2016}
Klauser C, Bigault T, B{\"o}ni P, Courtois P, Devishvili A, Rebrova N,
  Schneider M, Soldner T. 2016  Depolarization in polarizing supermirrors. {\em
  Nuclear Instruments and Methods in Physics Research Section A: Accelerators,
  Spectrometers, Detectors and Associated Equipment} \textbf{840}, 181 -- 185.

\end{thebibliography}
\end{document}